\documentclass[american,english]{IEEEtran}
\usepackage[T1]{fontenc}
\usepackage[latin9]{inputenc}
\usepackage{color}
\usepackage{mathtools}
\usepackage{amsmath}
\usepackage{amsthm}
\usepackage{amssymb}
\usepackage{cancel}
\usepackage{mathdots}
\usepackage{stmaryrd}
\usepackage{stackrel}
\usepackage{graphicx}
\PassOptionsToPackage{version=3}{mhchem}
\usepackage{mhchem}
\usepackage{esint}

\makeatletter
\theoremstyle{plain}
\newtheorem{thm}{\protect\theoremname}
\theoremstyle{plain}
\newtheorem{lem}[thm]{\protect\lemmaname}

\usepackage{colortbl}
\usepackage{cite}
\usepackage{stfloats}
\usepackage[margin=8pt,font=footnotesize,labelfont=md,labelsep=colon]{caption}

\usepackage{multicol}
\usepackage{float}
\usepackage{lipsum}
\@ifundefined{showcaptionsetup}{}{%
 \PassOptionsToPackage{caption=false}{subfig}}
\usepackage{subfig}
\makeatother

\usepackage{babel}
\addto\captionsamerican{\renewcommand{\lemmaname}{Lemma}}
\addto\captionsamerican{\renewcommand{\theoremname}{Theorem}}
\addto\captionsenglish{\renewcommand{\lemmaname}{Lemma}}
\addto\captionsenglish{\renewcommand{\theoremname}{Theorem}}
\providecommand{\lemmaname}{Lemma}
\providecommand{\theoremname}{Theorem}

\begin{document}
\title{On the Performance of Non-Orthogonal Multiple Access over Composite
Fading\textcolor{black}{{} Channels}}
\author{\textcolor{black}{\normalsize{}Khaled Rabie, Abubakar U. Makarfi,
}\textit{\textcolor{black}{\normalsize{}Member, IEEE}}\textcolor{black}{\normalsize{},
Rupak Kharel, }\textit{\textcolor{black}{\normalsize{}Senior Member,
IEEE}}\textcolor{black}{\normalsize{}, Osamah. S. Badarneh, }\textit{\textcolor{black}{\normalsize{}Member,
IEEE}}\textcolor{black}{\normalsize{}, Bamidele Adebisi, }\textit{\textcolor{black}{\normalsize{}Senior
Member, }}\textcolor{black}{\normalsize{}Xingwang Li}\textit{\textcolor{black}{\normalsize{},
Senior Member, IEEE }}\textcolor{black}{\normalsize{}and Zhiguo Ding}\textit{\textcolor{black}{\normalsize{},
Fellow, IEEE}}\\
\foreignlanguage{american}{\textcolor{black}{}}\thanks{\textcolor{black}{K. M. Rabie, A. U. Makarfi, R. Kharel and B. Adebisi
are with the Faculty of Science and Engineering, Manchester Metropolitan
University, UK, M1 6BH (e-mails: \{k.rabie; a.makarfi; r.kharel; b.adebisi\}@mmu.ac.uk).
}\protect \\
\textcolor{black}{O. S. Badarneh is with the Electrical and Communication
Engineering Department, German-Jordanian University, Jordan (e-mail:
Osamah.Badarneh@gju.edu.jo).}\protect \\
\textcolor{black}{X. Li is with }\textcolor{black}{\small{}school
of Physical and Electronics Engineering, Henan }\textcolor{black}{Polytechnic
University (HPU), China (e-mail: lixingwang@hpu.edu.cn).}\protect \\
\textcolor{black}{Z. Ding is with the School of Electrical and Electronic
Engineering, The University of Manchester, Manchester M13 9PL, U.K.
(e-mail: zhiguo.ding@manchester.ac.uk).}}}

\maketitle
\selectlanguage{american}%
\textcolor{black}{\thispagestyle{empty}}
\selectlanguage{english}%
\begin{abstract}
This paper analyzes the performance of a cooperative relaying non-orthogonal
multiple access (NOMA) network over Fisher-Snedecor $\mathcal{F}$
composite fading channels. Specifically, one base station (BS) is
assumed to communicate with two receiving mobile users with one acting
also as a decode-and-forward (DF) relay. To highlight the achievable
performance gains of the NOMA scheme, conventional relaying with orthogonal
multiple access (OMA) is also analyzed. For the two systems under
consideration, we derive novel exact closed-form expressions for the
ergodic capacity along with the corresponding asymptotic representations.
The derived expressions are then used to assess the influence of various
system parameters, including the fading and shadowing parameters,
on the performance of both the NOMA and OMA systems. Monte-Carlo simulation
are provided throughout to verify the accuracy of our analysis. Results
reveal that the NOMA system can considerably outperform the OMA approach
when the power allocation factor is carefully selected. It is also
shown that as the fading and/or shadowing parameters are increased,
the ergodic capacity performance enhances. 
\end{abstract}

\begin{IEEEkeywords}
\textcolor{black}{Conventional cooperative relaying (CCR), }Composite
fading, decode-and-forward (DF), Fisher-Snedecor $\mathcal{F}$ \textcolor{black}{model,
non-orthogonal multiple access (NOMA). }
\end{IEEEkeywords}

\section{\textcolor{black}{Introduction}}

\IEEEPARstart{N}{on}-orthogonal multiple access (NOMA) has recently
been proposed as a promising solution to overcome many challenges
facing the development of the fifth generation (5G) mobile communication
networks \cite{NOMAmain,NOMADing14,NOMAmain2,CoopNomaLi2}. In particular,
NOMA offers better spectral efficiency, balanced user fairness and
reduced access latency, which are crucial for the realization of numerous
Internet-of-things (IoT) applications. Unlike orthogonal multiple
access (OMA) systems such as frequency-division multiple access (FDMA)
and time-division multiple access (TDMA) where orthogonality is maintained
in frequency/time domain, NOMA can share the entire time and frequency
resources for all served users by superimposing their signals only
with different power levels \cite{DingMag17,NOMA_Yang17,NOMA_Yang172,RabieNOMA1}. 

Numerous studies have recently appeared in the literature analyzing
the performance of NOMA-based networks in various wireless communication
scenarios. For instance, multiple-input multiple-output (MIMO) techniques
were combined with NOMA to provide additional spatial degrees of freedom
\cite{NOMAmimo,NOMAmimo2}. NOMA with energy-harvesting (EH) capabilities
and various EH protocols were explored in \cite{NOMA_EH1,NOMA_EH2,NOMA_EH3,SWIPTcoopNomaLi,LiSWIPTnomaWeibull}.
In addition, physical layer security of NOMA-based networks was considered
in a number of recent works \cite{NOMA_PLSan,NOMAplsHe,NOMAplsLiu17,PLS_NOMA_Li}.
Particularly, however, NOMA in cooperative relaying wireless networks,
based on amplify-and-decode (AF) and decode-and-forward (DF) relaying
protocols, has recently attracted significant research attention,
see e.g., \cite{NOMA_RDing15,NOMA_RKim15,NOMA_MultiR_Ding16,NOMA_MultiR_Deng17,NOMArelays,PLS_NOMA_Li,SWIPTcoopNomaLi,IQnomaTian,HSUAVNOMA_Li}
and the references therein. More specifically, the authors in \cite{NOMA_RDing15}
were the first to investigate user cooperation in NOMA systems where
users with stronger channel conditions were assigned to assist the
other users based on DF relaying. In \cite{NOMA_RKim15}, the authors
studied the performance of a cooperative NOMA network with a single
dedicated relay helping multiple users using AF and DF schemes. Since
more relays provide higher diversity gains, the performance of multi-relay
NOMA networks was investigated in \cite{NOMA_MultiR_Ding16}. Furthermore,
the authors in \cite{NOMA_MultiR_Deng17} proposed a joint user and
relay selection for cooperative NOMA systems in which multiple users
communicate with two destinations through multiple AF relays. In addition,
full-duplex relaying in cooperative NOMA systems were studied in \cite{NOMA_FDYue18,NOMA_FDZhang17,NOMA_FDZhang17a,LiFDcoopNOMA}.

Indeed, all the aforementioned studies have indicated the significance
of channel fading characteristics on the communication performance
of NOMA-based system\textcolor{black}{s. }To the authors' best knowledge,
none of the studies above studied the performance of NOMA networks
over composite fading channels which consider concurrently the effects
of both multipath fading and shadowing. To fill this gap, this paper
is dedicated to analyzing the performance of cooperative relaying
NOMA systems in composite fading channels; more specifically, we adopt
the recently introduced Fisher-Snedecor $\mathcal{F}$ composite model
\cite{FisherModel17}, in which it is assumed that scattered multipath
has a Nakagami-$m$ distribution whereas the root-mean-square signal
follows an inverse Nakagami-$m$ distribution. \textcolor{black}{The
motivation for adopting }this channel model\textcolor{black}{{} is threefold.
Firstly, the Fisher-Snedecor $\mathcal{F}$ distribution is accurate
and has better tractability compared to other models. Secondly, the
$\mathcal{F}$ fading model has been shown to perform better than
other composite fading distributions such as generalized-$K$ model
in terms of modeling accuracy and computational complexity \cite{Osamah18WiMob}.
Finally, the Fisher-Snedecor $\mathcal{F}$ model includes several
fading distributions as special cases, such as Nakagami-$m$ }($m_{s}\rightarrow\infty$)\textcolor{black}{{}
and Rayleigh} ($m_{s}\rightarrow\infty,$ $m=1$)\textcolor{black}{,
where $m_{s}$ and $m$ are the shadowing and fading severity parameters,
respectively.}

In this work, we consider a NOMA-based cooperative relaying scenario
with one base station (BS) communicating directly with two users in
the presence of Fisher-Snedecor $\mathcal{F}$ composite fading. The
major contributions of this study are summarized as follows. 
\begin{itemize}
\item \textcolor{black}{We derive novel, closed-form expressions for the
probability density function (PDF) and cumulative distribution function
(CDF) of the minimum of two Fisher-Snedecor $\mathcal{F}$ random
variables (RVs) in terms of the Meijer\textquoteright s G-function.
These fundamental statistics are essential for the computation of
the communications performance metrics of interest.}
\item \textcolor{black}{The exact ergodic capacity expressions of a two-hop
DF cooperative system with NOMA and OMA schemes over Fisher-Snedecor
$\mathcal{F}$ composite fading channels are derived by utilizing
the obtained PDF and CDF expressions. }
\item \textcolor{black}{The corresponding asymptotic behavior of the ergodic
capacity in the high signal-to-noise ratio (SNR) regime is investigated
for the two systems under study. }
\item \textcolor{black}{Through the derived expressions, we examine the
impact of various system and fading parameters on the capac}ity performance.
This also allowed us to compare the performance of both the cooperative
NOMA and OMA systems. The correctness of the derived analysis is demonstrated
by means of equivalent results obtained via Monte-Carlo simulations.
\end{itemize}
It is worthwhile mentioning that the ergodic capacity expressions
are given in terms of special functions such as univariate and multivariate
Meijer G-functions, which have recently been extensively used in the
literature \cite{Balti18,Lei18,GalymAccess18,An16}. Furthermore,
although some expressions are expressed in terms of infinite series,
they converge rapidly with rather few terms for obtaining a given
numerical accuracy. 

The rest of this paper is organized as follows. Section \ref{sec:System-Model}
describes the system and channel models considered in this work. In
Section \ref{sec:two-fisher-min}, we characterize the minimum of
two Fisher-Snedecor $\mathcal{F}$ RVs and present new closed-form
expressions for the corresponding PDF and CDF. In Sections \ref{sec:Cooperative-Relaying-NOMA}
and \ref{sec:Cooperative-OMA}, we derive closed-form analytical expressions
for the ergodic capacity for both the NOMA and conventional cooperative
relaying systems over the Fisher-Snedecor $\mathcal{F}$ composite
fading channel, respectively. Section \ref{sec:Numerical-Examples}
presents numerical examples of the derived expressions with some insights
and discussions of results. Finally, Section \ref{sec:Conclusion}
concludes the paper with a summary of the main results.

The following notations are used in this paper. $f_{X}\left(\cdot\right)$,
$F_{X}\left(\cdot\right)$ and $\bar{F}_{X}\left(\cdot\right)$ denote
the PDF, CDF and the complementary CDF (CCDF) of the random variable
(RV) $X$. $\mathbb{E}\left[\cdot\right]$, $\bigl|\cdot\bigr|$,
$\sum$ and $\prod$ are the expectation, absolute value, summation
and product operators. $\textrm{min}\left\{ \cdot\right\} $ and $\textrm{ln}\left(\cdot\right)$
represent the minimum argument and the natural logarithm, respectively.
The symbol $\left(x\right)_{k}$ is the Pochhammer symbol, $B\left(\cdot,\cdot\right)$
is the Beta function \cite[eq. (8.4.2.5)]{book2}, $\Gamma\left[\cdot\right]$
is the Gamma function \cite[eq. 8.310.1]{BookGrad2000}, $\psi^{\left(0\right)}\left(\cdot\right)$
is the digamma function, $\textrm{G}_{p,q}^{n,m}\left[\cdot|\cdot\right]$
is the Meijer G-function \cite[eq. 9.301]{BookGrad2000}, $_{2}F_{1}\left(\cdot,\cdot;\cdot;\cdot\right)$
is the Gauss hypergeometric function \cite[Eq. (9.100)]{book2} and
$F_{1}\left(\cdot;\cdot,\cdot;\cdot;\cdot,\cdot\right)$ is the Appell
two-variables hypergeometric function \cite[Eq. (9.14.1; 9.210)]{book2}.

\section{System Model\textcolor{black}{\label{sec:System-Model}}}

Let us consider a wireless network illustrated in Fig. 1, which consists
of three nodes: a BS, a far user (D) and a near user (R), with the
latter acting also as a decode-and-forward (DF) relay operating in
half duplex mode. It is assumed that all nodes are equipped with a
single antenna and that there exists a direct link between the BS
and the far user. The S-to-R, R-to-D and S-to-D channel coefficients
are denoted by $h_{\textrm{sr}}$, $h_{\textrm{rd}}$ and $h_{\textrm{sd}}$,
respectively, all of which experience independent quasi-static Fisher-Snedecor
$\mathcal{F}$ fading. The corresponding instantaneous SNRs of these
links are denoted by $\gamma_{i}=\bigl|h_{i}\bigr|^{2},$ where $i\in\left\{ \textrm{sr},\textrm{rd},\textrm{sd}\right\} ,$
with the PDF in (\ref{eq:CDFXY}).
\begin{figure}
\begin{centering}
\includegraphics[bb=140bp 520bp 400bp 670bp,clip,scale=0.85]{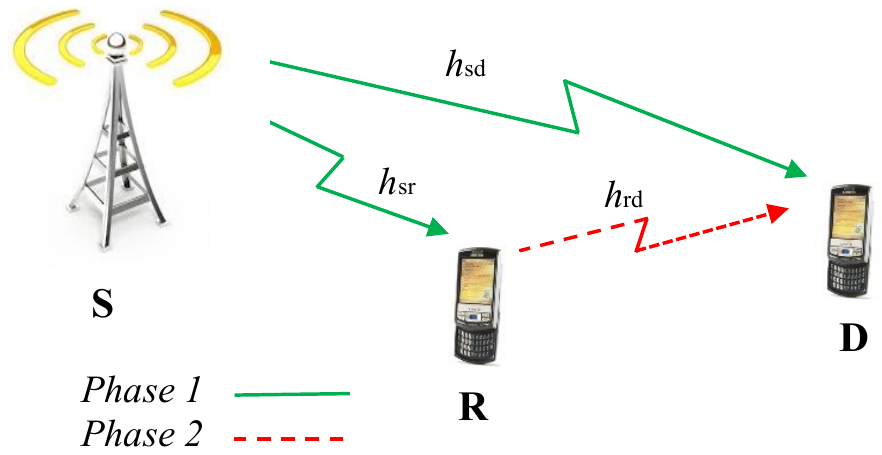}
\par\end{centering}
\caption{The considered NOMA system model with a BS and two users over Fisher-Snedecor
$\mathcal{F}$ composite fading channels.}
\end{figure}

\textcolor{black}{The considered NOMA system requires two time slots
to convey information to the two users. In the first time slot, the
BS transmits the signal $\sum_{k\in\left\{ 1,2\right\} }$ $\sqrt{a_{k}P_{s}}s_{k}$
to R and D, where $P_{s}$ is the BS transmit power whereas $a_{1}$
and $a_{2}$ are the power allocation factors for D and R, respectively.
Note that $a_{1}+a_{2}=1$ and $a_{1}>a_{2}$. With this in mind,
the received signals at R and D can be expressed as }

\textcolor{black}{
\begin{equation}
y_{i}\left(t\right)=h_{i}\sum_{k\in\left\{ 1,2\right\} }\sqrt{\frac{a_{k}P_{s}}{d_{i}^{\beta_{i}}}}s_{k}\left(t\right)+n\left(t\right),
\end{equation}
}

\noindent \textcolor{black}{where $i\in\left\{ \textrm{sr},\textrm{sd}\right\} $
and $n\left(t\right)$ is the noise at R and D which is assumed to
be Additive white Gaussian noise (AWGN) with zero mean and variance
$\sigma^{2}$, while $\beta_{i}$ and $d_{i}$ are the path-loss exponents
and distances from S to R and D. The relay during the first time slot
decodes $s_{1}$ and treats $s_{2}$ as noise, and to obtain the latter,
the former is canceled using successive interference cancellation
(SIC). In this respect, the corresponding SNRs of $s_{1}$ and $s_{2}$
during the first time slot can be written as }

\begin{equation}
\gamma_{\textrm{sr}}^{\left(1\right)}=\frac{a_{1}P_{s}\bigl|h_{\textrm{sr}}\bigr|^{2}}{a_{2}P_{s}\bigl|h_{\textrm{sr}}\bigr|^{2}+\sigma^{2}},\,\textrm{and}\label{eq:snrSR1}
\end{equation}

\textcolor{black}{
\begin{equation}
\gamma_{\textrm{sr}}^{\left(2\right)}=\frac{a_{2}P_{s}\bigl|h_{\textrm{sr}}\bigr|^{2}}{\sigma^{2}}.\hfill\label{eq:snrSR2}
\end{equation}
}

\textcolor{black}{At D, because the symbol $s_{2}$ will be treated
as noise during the first time slot, the corresponding SNR is expressed
as }

\textcolor{black}{
\begin{equation}
\gamma_{\textrm{sd}}=\frac{a_{1}P_{s}\bigl|h_{\textrm{sd}}\bigr|^{2}}{a_{2}P_{s}\bigl|h_{\textrm{sd}}\bigr|^{2}+\sigma^{2}}.\label{eq:snrSD}
\end{equation}
}

\textcolor{black}{The relay in the second time slot, assuming that
it successfully decoded $s_{2}$, will forward this symbol to D. Therefore,
the received signal at D in the second time slot is $y_{\textrm{rd}}\left(t\right)=\sqrt{P_{r}}h_{\textrm{rd}}s_{2}\left(t\right)+n\left(t\right)$
and hence the corresponding SNR is }

\begin{equation}
\gamma_{\textrm{rd}}=\frac{P_{r}\bigl|h_{\textrm{rd}}\bigr|^{2}}{\sigma^{2}},\label{eq:snrRD}
\end{equation}

\noindent where $P_{r}$ is the relay transmit power.

\section{\textcolor{black}{Characterization of the Minimum of Two }Fisher-Snedecor
$\mathcal{F}$\textcolor{black}{{} RVs\label{sec:two-fisher-min} }}

In this section, we derived closed-form expressions for the CDF and
PDF of the minimum of two Fisher-Snedecor $\mathcal{F}$ variates,
which are essential to carry out our analysis in this work. To begin
with, let us define two RVs, $\gamma_{x}$ and $\gamma_{y}$, that
follow the Fisher-Snedecor $\mathcal{F}$\textcolor{black}{{} distribution
with PDF} \cite{FisherModel17}

\textcolor{black}{\small{}
\begin{align}
f_{\gamma_{i}}\left(\gamma_{i}\right) & =\frac{m_{i}^{m_{i}}\left(m_{s_{i}}-1\right)^{m_{s_{i}}}\bar{\gamma}^{m_{s_{i}}}\gamma_{i}^{m_{i}-1}}{B\left(m_{i},\,m_{s_{i}}\right)\left(m\gamma+\left(m_{s_{i}}-1\right)\bar{\gamma}_{i}\right)^{m_{i}+m_{s_{i}}}},\nonumber \\
 & \stackrel{\left(a\right)}{=}\frac{1}{\Gamma\left(m_{i}\right)\Gamma\left(m_{s_{i}}\right)}\gamma_{i}^{-1}\textrm{G}_{1,1}^{1,1}\left[\Lambda_{i}\,\gamma_{i}\Biggl|\negthickspace\begin{array}{c}
1-m_{s_{i}}\\
m_{i}
\end{array}\negthickspace\right],m_{s_{i}}>1\label{eq:pdfXY}
\end{align}
}{\small\par}

\noindent while $i\in\left\{ x,y\right\} $, $\bar{\gamma}_{i}=\mathbb{E}\left[\gamma_{i}\right]$
is the mean power and $\Lambda_{i}=\frac{m_{i}}{\left(m_{s_{i}}-1\right)\bar{\gamma}_{i}}$,
$m_{i}$ and $m_{s_{i}}$ represent respectively the multipath severity
and shadowing parameters of the $i$-th RV, respectively. Note that
$\left(a\right)$ in (\ref{eq:pdfXY}) is obtained with the aid of
\cite[Eq. (8.4.2.5)]{bookV3} and \cite[Eq. (8.2.2.15)]{bookV3}. 

Integrating (\ref{eq:pdfXY}) with respect to $\gamma_{i}$, with
the help of \cite[Eq. (26)]{Adamchik90}, we can express the CDF of
$\gamma_{i}$ as 
\begin{equation}
F_{\gamma_{i}}\left(\gamma_{i}\right)=\frac{1}{\Gamma\left(m_{i}\right)\Gamma\left(m_{s_{i}}\right)}\,\textrm{G}_{2,2}^{1,2}\left[\Lambda_{i}\gamma_{i}\Biggl|\negthickspace\begin{array}{c}
1-m_{s_{i}},\,1\\
m_{i},\,0
\end{array}\negthickspace\right].\label{eq:CDFXY}
\end{equation}

Now, let us define the RV 

\begin{equation}
\gamma_{z}\triangleq\textrm{min}\left\{ \gamma_{x},\,\gamma_{y}\right\} ,
\end{equation}

\noindent where $\gamma_{x}$ and $\gamma_{y}$ are independent and
non-identically distributed Fisher-Snedecor $\mathcal{F}$ RVs. 

With this in mind, the CDF of $\gamma_{z}$ can be obtained as follows

\begin{equation}
F_{\gamma_{z}}\left(u\right)=1-\bar{F}_{\gamma_{x}}\left(u\right)\bar{F}_{\gamma_{y}}\left(u\right),\label{eq:CDFZ0}
\end{equation}

\noindent where $\bar{F}_{\gamma_{i}}\left(u\right)$, is the CCDF
of $\gamma_{i}$ which is related to its CDF as $\bar{F}_{\mathcal{X}_{i}}\left(\cdot\right)=1-F_{\mathcal{X}_{i}}\left(\cdot\right)$.
Since $\gamma_{x}$ and $\gamma_{y}$ have Fisher-Snedecor $\mathcal{F}$
distribution, their CCDFs can be obtained straightforwardly from (\ref{eq:CDFXY})
with the appropriate notations change. In light of this, using (\ref{eq:CDFXY})
and (\ref{eq:CDFZ0})\textcolor{black}{, along performing some straightforward
manipulations, th}e CDF of $\gamma_{z}$ can be expressed as 

\begin{align}
F_{\gamma_{z}}\left(u\right) & =\underset{i\in\left\{ x,y\right\} }{\sum}\Theta_{i}\,\textrm{G}_{2,2}^{1,2}\left[\,u\,\Biggl|\negthickspace\begin{array}{c}
1-m_{s_{i}},\,1\\
m_{i},\,0
\end{array}\negthickspace\right]\nonumber \\
 & \qquad-\underset{i\in\left\{ x,y\right\} }{\prod}\Theta_{i}\,G_{2,2}^{1,2}\left[\,u\,\Biggl|\negthickspace\begin{array}{c}
1-m_{s_{i}},\,1\\
m_{i},\,0
\end{array}\negthickspace\right],\label{eq:CDFZ}
\end{align}

\noindent where $\Theta_{i}=1/\left\{ \Gamma\left(m_{i}\right)\Gamma\left(m_{s_{i}}\right)\right\} .$ 

Taking the derivative of $F_{\gamma_{z}}\left(u\right)$, and invoking
\cite[Eq. (8.2.2.30)]{bookV3}, we obtain the PDF of $\gamma_{z}$
as follows 

\begin{align}
\negthickspace\negthickspace\negthickspace f_{\gamma_{z}}\left(z\right)=\underset{i\in\left\{ x,y\right\} }{\sum}\Theta_{i}\,G_{3,3}^{1,3}\left[\,z\,\Biggl|\negthickspace\begin{array}{c}
-1,\,-m_{s_{i}},\,0\\
m_{i}-1,\,0,\,-1
\end{array}\negthickspace\right]\qquad\nonumber \\
-\underset{i\in\left\{ x,y\right\} }{\prod}\Theta_{i}\,G_{2,2}^{1,2}\left[\,z\,\Biggl|\negthickspace\begin{array}{c}
1-m_{s_{i}},\,1\\
m_{i},\,0
\end{array}\negthickspace\right] & ,\label{eq:PDFZ}
\end{align}

To the authors' best knowledge, (\ref{eq:CDFZ}) and (\ref{eq:PDFZ})
are new. It is important to mention that, with the help of \cite[Eq. (8.2.2.15)]{bookV3},
\cite[Eq. (8.4.2.5)]{bookV3} and \cite[Eq. (17)]{Adamchik90}, the
PDF in (\ref{eq:PDFZ}) can also be written in the following form 

\begin{align}
f_{\gamma_{z}}\left(z\right)=\underset{i\in\left\{ x,y\right\} }{\sum}\frac{z^{m_{i}-1}\left(1+z\right)^{m_{i}+m_{s_{i}}}}{B\left(m_{i},m_{s_{i}}\right)}-\underset{i\in\left\{ x,y\right\} }{\prod}\frac{z^{m_{i}}}{m_{i}}\nonumber \\
\times\frac{1}{B\left(m_{i},m_{s_{i}}\right)}{}_{2}F_{1}\left(m_{i},m_{i}+m_{s_{i}};m_{i}+1;-z\right).\label{eq:PDFZ2}
\end{align}

\section{\label{sec:Cooperative-Relaying-NOMA}Cooperative NOMA over $\mathcal{F}$
Fading Channels }

We analyze in this section the exact and asymptotic ergodic sum rate
performance of the cooperative NOMA system over the Fisher-Snedecor
$\mathcal{F}$ composite fading channels. 

\subsection{Exact Performance Analysis }

\textcolor{black}{The ergodic sum rate of the cooperative NOMA system
with DF relaying, $\bar{C}_{\textrm{NOMA}}$, consists of the two
rates associated with the symbols $s_{1}$ and $s_{2}$, denoted respectively
as $\bar{C}_{1}$ and $\bar{C}_{2}$. That is }

\begin{equation}
\bar{C}_{\textrm{NOMA}}=\bar{C}_{1}+\bar{C}_{2}\label{eq:Cnoma}
\end{equation}

\textcolor{black}{To simplify our notations and without loss of generality,
we assume from now onward that $P=P_{\textrm{s}}=P_{\textrm{r}}$
and $\bar{\gamma}=P/\sigma^{2}$. The achievable rate associated with
the first symbol can be calculated as}
\begin{equation}
C_{1}=\frac{1}{2}\,\textrm{min}\left\{ \textrm{log}_{2}\left(1+\gamma_{\textrm{sr}}^{(1)}\right),\,\textrm{log}_{2}\left(1+\gamma_{\textrm{sd}}\right)\right\} ,\label{eq:C1}
\end{equation}

\noindent which, using (\ref{eq:snrSR1}) and (\ref{eq:snrSD}), can
be rewritten as $C_{1}=C_{1,1}-C_{1,2}$, where

\noindent 
\begin{equation}
C_{1,1}=\frac{1}{2}\,\textrm{log}_{2}\left(1+\textrm{min}\left\{ \bigl|h_{\textrm{sr}}\bigr|^{2},\,\bigl|h_{\textrm{sd}}\bigr|^{2}\right\} \bar{\gamma}\right)\,\textrm{and}\label{eq:C11a}
\end{equation}
\begin{align}
C_{1,2} & =\frac{1}{2}\,\textrm{log}_{2}\left(1+\textrm{min}\left\{ \bigl|h_{\textrm{sr}}\bigr|^{2},\,\bigl|h_{\textrm{sd}}\bigr|^{2}\right\} a_{2}\bar{\gamma}\right).\label{eq:C12a}
\end{align}

On the other hand, the achievable rate associated with the second
symbol can be determined as
\begin{align}
C_{2} & =\frac{1}{2}\,\textrm{min}\left\{ \textrm{log}_{2}\left(1+\gamma_{\textrm{sr}}^{(2)}\right),\,\textrm{log}_{2}\left(1+\gamma_{\textrm{rd}}\right)\right\} \nonumber \\
 & =\frac{1}{2}\,\textrm{log}_{2}\left(1+\textrm{min}\left\{ \bigl|h_{\textrm{sr}}\bigr|^{2}a_{2},\,\bigl|h_{\textrm{rd}}\bigr|^{2}\right\} \bar{\gamma}\right).\label{eq:C2main}
\end{align}

Letting $\mathcal{X}=\textrm{min}\left\{ \mathcal{X}_{\textrm{sr}},\,\mathcal{X}_{\textrm{sd}}\right\} $,
where $\mathcal{X}_{\textrm{sr}}=\bar{\gamma}\bigl|h_{\textrm{sr}}\bigr|^{2}$
and $\mathcal{X}_{\textrm{sd}}=\bar{\gamma}\bigl|h_{\textrm{sd}}\bigr|^{2}$,
and using (\ref{eq:CDFZ}) with the appropriate change of variables,
we can obtain the CDF of $\mathcal{X}$ as

\begin{align}
F_{\mathcal{X}}\left(u\right) & =\underset{i\in\left\{ \textrm{sr},\textrm{sd}\right\} }{\sum}\Theta_{i}\,\textrm{G}_{2,2}^{1,2}\left[\Lambda_{i}\,u\Biggl|\negthickspace\begin{array}{c}
1-m_{s_{i}},\,1\\
m_{i},\,0
\end{array}\negthickspace\right]\nonumber \\
 & \qquad-\underset{i\in\left\{ \textrm{sr},\textrm{sd}\right\} }{\prod}\Theta_{i}\,G_{2,2}^{1,2}\left[\Lambda_{i}\,u\Biggl|\negthickspace\begin{array}{c}
1-m_{s_{i}},\,1\\
m_{i},\,0
\end{array}\negthickspace\right],\label{eq:CDFX2}
\end{align}

\noindent where $\Theta_{i}=1/\left\{ \Gamma\left(m_{i}\right)\Gamma\left(m_{s_{i}}\right)\right\} .$ 

Taking the derivative of $F_{\mathcal{X}}\left(u\right)$, and with
the aid of \cite[Eq. (8.2.2.30)]{bookV3}, we obtain the PDF of $\mathcal{X}$
as follows 

\begin{align}
\negthickspace\negthickspace\negthickspace f_{\mathcal{X}}\left(z\right)=\underset{i\in\left\{ \textrm{sr},\textrm{sd}\right\} }{\sum}\Lambda_{i}\Theta_{i}\,G_{3,3}^{1,3}\left[\Lambda_{i}\,z\Biggl|\negthickspace\begin{array}{c}
-1,\,-m_{s_{i}},\,0\\
m_{i}-1,\,0,\,-1
\end{array}\negthickspace\right]\qquad\nonumber \\
-\underset{i\in\left\{ \textrm{sr},\textrm{sd}\right\} }{\prod}\Theta_{i}\,G_{2,2}^{1,2}\left[\Lambda_{i}\,z\Biggl|\negthickspace\begin{array}{c}
1-m_{s_{i}},\,1\\
m_{i},\,0
\end{array}\negthickspace\right] & .\label{eq:PDFX}
\end{align}

To this end, the average rate $\bar{C}_{1,1}$ can be obtained as
$\bar{C}_{1,1}=\stackrel[0]{\infty}{\int}f_{\mathcal{X}}\left(z\right)$
$\textrm{ln}\left(1+z\right)\textrm{d}z$. Using (\ref{eq:PDFX})
while expressing the natural logarithmic function in terms of the
Meijer G-function, i.e., $\textrm{ln}\left(1+z\right)=G_{2,2}^{1,2}\biggl[z|\!\begin{array}{c}
1,1\\
1,0
\end{array}\!\biggr]$, \cite[Eq. (11)]{Adamchik90}, we can express $\bar{C}_{1,1}$ as
in (\ref{eq:EC11I}), shown at the top of this page. 
\begin{figure*}[t]
\begin{eqnarray}
\bar{C}_{1,1}=\underset{i\in\left\{ \textrm{sr},\textrm{sd}\right\} }{\sum}\frac{\Lambda_{i}}{\Gamma\left(m_{i}\right)\Gamma\left(m_{s_{i}}\right)}\underset{I_{1}}{\underbrace{\stackrel[0]{\infty}{\int}G_{3,3}^{1,3}\left[\Lambda_{i}\,z\Biggl|\negthickspace\begin{array}{c}
-1,\,-m_{s_{i}},\,0\\
m_{i}-1,\,0,\,-1
\end{array}\negthickspace\right]\textrm{G}_{2,2}^{1,2}\left[z\biggl|\!\begin{array}{c}
1,1\\
1,0
\end{array}\!\right]\textrm{d}z}}-\underset{i\in\left\{ \textrm{sr},\textrm{sd}\right\} }{\prod}\frac{1}{\Gamma\left(m_{i}\right)\Gamma\left(m_{s_{i}}\right)}\quad\quad\nonumber \\
\negthickspace\negthickspace\negthickspace\negthickspace\negthickspace\negthickspace\negthickspace\negthickspace\negthickspace\negthickspace\negthickspace\negthickspace\negthickspace\negthickspace\negthickspace\negthickspace\negthickspace\negthickspace\negthickspace\negthickspace\negthickspace\negthickspace\negthickspace\negthickspace\negthickspace\negthickspace\negthickspace\negthickspace\negthickspace\negthickspace\negthickspace\negthickspace\negthickspace\times\underset{I_{2}}{\underbrace{\stackrel[0]{\infty}{\int}\textrm{G}_{2,2}^{1,2}\left[\Lambda_{\textrm{sr}}\,z\Biggl|\negthickspace\begin{array}{c}
1-m_{s_{\textrm{sr}}},\,1\\
m_{\textrm{sr}},\,0
\end{array}\negthickspace\right]G_{2,2}^{1,2}\left[\Lambda_{\textrm{sd}}\,z\Biggl|\negthickspace\begin{array}{c}
1-m_{s_{\textrm{sd}}},\,1\\
m_{\textrm{sd}},\,0
\end{array}\negthickspace\right]\textrm{G}_{2,2}^{1,2}\left[z|\!\begin{array}{c}
1,1\\
1,0
\end{array}\!\right]\textrm{d}z}}.\label{eq:EC11I}
\end{eqnarray}

\selectlanguage{american}%
\centering{}\rule[0.5ex]{2.03\columnwidth}{0.8pt}\selectlanguage{english}%
\end{figure*}
 With the help of \cite[Eq. (7.811.1)]{book2}, the integral $I_{1}$
in (\ref{eq:EC11I}) can be solved as 
\begin{equation}
I_{1}=\textrm{G}_{5,5}^{4,3}\left[\frac{1}{\Lambda_{i}}\Biggl|\negthickspace\begin{array}{c}
1-m_{i},\,1,\,1,\,0,\,1\\
1,\,m_{s_{i}},\,0,\,1,\,0
\end{array}\negthickspace\right],\label{eq:I1}
\end{equation}

\noindent whereas the second integral $I_{2}$ is solved with the
aid of \cite{3Gfunctions.} as follows 

{\small{}
\begin{equation}
I_{2}=\textrm{G}_{2,2:2,2:3,3}^{2,1:1,2:1,3}\left[\Lambda_{i},\Lambda_{l}\Biggl|\negthickspace\begin{array}{c}
-1,0\\
-1,-1
\end{array}\negthickspace\Biggl|\negthickspace\begin{array}{c}
1-m_{s_{i}},1\\
m_{i},0
\end{array}\negthickspace\Biggl|\negthickspace\begin{array}{c}
-1,-m_{s_{l}},0\\
m_{l}-1,0,-1
\end{array}\negthickspace\right],\label{eq:I2}
\end{equation}
}where $\textrm{ G}_{p_{1},q_{1}:p_{2},q_{2}:p_{3},q_{3}}^{m_{1},n_{1}:m_{2},n_{2}:m_{3},n_{3}}\bigl[\cdot\bigr]$
is the bivariate Meijer G-function \cite{bookBiGfun}. 

Substituting (\ref{eq:I1}) and (\ref{eq:I2}) into (\ref{eq:EC11I})
yields a closed-form expression for average capacity $\bar{C}_{1,1}$
given in (\ref{eq:EC11Final}), shown at the top of the next page.
\begin{gather}
\bar{C}_{1,1}=\underset{i\in\left\{ \textrm{sr},\textrm{sd}\right\} }{\sum}\frac{1}{\Gamma\left(m_{i}\right)\Gamma\left(m_{s_{i}}\right)}\,\textrm{G}_{5,5}^{4,3}\left[\frac{1}{\Lambda_{i}}\Biggl|\begin{array}{c}
1-m_{i},\,1,\,1,\,0,\,1\\
1,\,m_{s_{i}},\,0,\,1,\,0
\end{array}\right]\nonumber \\
-\underset{\underset{i\neq l}{i,l\in\left\{ \textrm{sr},\textrm{sd}\right\} }}{\sum}\frac{\Lambda_{l}}{\underset{j\in\left\{ \textrm{sr},\textrm{rd}\right\} }{\prod}\Gamma\left(m_{j}\right)\Gamma\left(m_{s_{j}}\right)}\times\nonumber \\
\textrm{ G}_{2,2:2,2:3,3}^{2,1:1,2:1,3}\left[\Lambda_{i},\,\Lambda_{l}\Biggl|\begin{array}{c}
-1,\,0\\
-1,\,-1
\end{array}\Biggl|\begin{array}{c}
1-m_{s_{i}},\,1\\
m_{i},\,0
\end{array}\Biggl|\begin{array}{c}
-1,-m_{s_{l}},\,0\\
m_{l}-1,\,0,\,-1
\end{array}\right]\label{eq:EC11Final}
\end{gather}

To find the average capacity $\bar{C}_{1,2}$, let $\mathcal{Y}=\textrm{min}\left\{ \mathcal{Y}_{\textrm{sr}},\,\mathcal{Y}_{\textrm{sd}}\right\} $,
where $\mathcal{Y}_{\textrm{sr}}=\bar{\gamma}a_{2}\bigl|h_{\textrm{sr}}\bigr|^{2}$
and $\mathcal{Y}_{\textrm{sd}}=\bar{\gamma}a_{2}\bigl|h_{\textrm{sd}}\bigr|^{2}$.
Following the same procedure used to derive $\bar{C}_{1,1}$, it is
straightforward to show that average capacity $\bar{C}_{1,2}$ can
be obtained in closed-form as in (\ref{eq:C12}), shown at the top
of the next page. The derivation is omitted for the sake of brevity.
\begin{figure*}[t]
\begin{eqnarray}
\bar{C}_{1,2}=\underset{i\in\left\{ \textrm{sr},\textrm{sd}\right\} }{\sum}\frac{1}{\Gamma\left(m_{i}\right)\Gamma\left(m_{s_{i}}\right)}\,\textrm{G}_{5,5}^{4,3}\left[\frac{a_{2}}{\Lambda_{i}}\Biggl|\negthickspace\begin{array}{c}
1-m_{i},\,1,\,1,\,0,\,1\\
1,\,m_{s_{i}},\,0,\,1,\,0
\end{array}\negthickspace\right]-\underset{\underset{i\neq k}{i,l\in\left\{ \textrm{sr},\textrm{sd}\right\} }}{\sum}\frac{\Lambda_{l}}{\underset{j\in\left\{ \textrm{sr},\textrm{rd}\right\} }{\prod}\Gamma\left(m_{j}\right)\Gamma\left(m_{s_{j}}\right)a_{2}}\quad\quad\quad\nonumber \\
\negthickspace\negthickspace\negthickspace\negthickspace\negthickspace\negthickspace\negthickspace\negthickspace\negthickspace\negthickspace\negthickspace\negthickspace\negthickspace\negthickspace\negthickspace\negthickspace\negthickspace\negthickspace\negthickspace\negthickspace\negthickspace\negthickspace\negthickspace\negthickspace\negthickspace\negthickspace\negthickspace\negthickspace\negthickspace\negthickspace\negthickspace\negthickspace\negthickspace\times\textrm{ G}_{2,2:2,2:3,3}^{2,1:1,2:1,3}\left[\frac{\Lambda_{i}}{a_{2}},\,\frac{\Lambda_{l}}{a_{2}}\Biggl|\negthickspace\begin{array}{c}
-1,\,0\\
-1,\,-1
\end{array}\negthickspace\Biggl|\negthickspace\begin{array}{c}
1-m_{s_{i}},\,1\\
m_{i},\,0
\end{array}\negthickspace\Biggl|\negthickspace\begin{array}{c}
-1,-m_{s_{l}},\,0\\
m_{l}-1,\,0,\,-1
\end{array}\negthickspace\right]\label{eq:C12}
\end{eqnarray}

\selectlanguage{american}%
\centering{}\rule[0.5ex]{2.03\columnwidth}{0.8pt}\selectlanguage{english}%
\end{figure*}

Next we derive the average capacity associated with the second symbol
$\bar{C}_{2}$. From (\ref{eq:C2main}), let $\mathcal{Z}=\textrm{min}\left\{ \mathcal{Z}_{\textrm{sr}},\,\mathcal{Z}_{\textrm{sd}}\right\} $,
where $\mathcal{Z}_{\textrm{sr}}=\bar{\gamma}a_{2}\bigl|h_{\textrm{sr}}\bigr|^{2}$
and $\mathcal{Z}_{\textrm{rd}}=\bar{\gamma}\bigl|h_{\textrm{rd}}\bigr|^{2}$
and following the same steps used to analyze $\mathcal{X}$ and $\mathcal{Y}$,
we can express $\bar{C}_{2}$ in closed-form as in (\ref{eq:C2}),
shown at the top of the next page. 
\begin{figure*}[t]
\begin{eqnarray}
\bar{C}_{2}=\frac{1}{\Gamma\left(m_{\textrm{sr}}\right)\Gamma\left(m_{s_{\textrm{sr}}}\right)}\,\textrm{G}_{5,5}^{4,3}\left[\frac{a_{2}}{\Lambda_{\textrm{sr}}}\Biggl|\negthickspace\begin{array}{c}
1-m_{\textrm{sr}},\,1,\,1,\,0,\,1\\
1,\,m_{s_{\textrm{sr}}},\,0,\,1,\,0
\end{array}\negthickspace\right]+\frac{1}{\Gamma\left(m_{\textrm{rd}}\right)\Gamma\left(m_{s_{\textrm{rd}}}\right)}\,\textrm{G}_{5,5}^{4,3}\left[\frac{1}{\Lambda_{\textrm{rd}}}\Biggl|\negthickspace\begin{array}{c}
1-m_{\textrm{rd}},\,1,\,1,\,0,\,1\\
1,\,m_{s_{\textrm{rd}}},\,0,\,1,\,0
\end{array}\negthickspace\right]\quad\quad\quad\nonumber \\
\negthickspace\negthickspace\negthickspace\negthickspace\negthickspace\negthickspace\negthickspace\negthickspace\negthickspace\negthickspace\negthickspace\negthickspace\negthickspace\negthickspace\negthickspace\negthickspace\negthickspace\negthickspace\negthickspace\negthickspace\negthickspace\negthickspace\negthickspace\negthickspace\negthickspace\negthickspace\negthickspace\negthickspace\negthickspace\negthickspace\negthickspace\negthickspace\negthickspace-\frac{1}{\underset{j\in\left\{ \textrm{sr},\textrm{rd}\right\} }{\prod}\Gamma\left(m_{j}\right)\Gamma\left(m_{s_{j}}\right)}\Biggl(\Lambda_{\textrm{rd}}\,\textrm{G}_{2,2:2,2:3,3}^{2,1:1,2:1,3}\left[\Lambda_{\textrm{rd}},\,\frac{\Lambda_{\textrm{sr}}}{a_{2}}\Biggl|\negthickspace\begin{array}{c}
-1,\,0\\
-1,\,-1
\end{array}\negthickspace\Biggl|\negthickspace\begin{array}{c}
1-m_{s_{\textrm{sr}}},\,1\\
m_{\textrm{sr}},\,0
\end{array}\negthickspace\Biggl|\negthickspace\begin{array}{c}
-1,-m_{s_{\textrm{rd}}},\,0\\
m_{\textrm{rd}}-1,\,0,\,-1
\end{array}\negthickspace\right]\nonumber \\
+\,\frac{\Lambda_{\textrm{sr}}}{a_{2}}\,\textrm{G}_{2,2:2,2:3,3}^{2,1:1,2:1,3}\left[\Lambda_{\textrm{rd}},\,\frac{\Lambda_{\textrm{sr}}}{a_{2}}\Biggl|\negthickspace\begin{array}{c}
-1,\,0\\
-1,\,-1
\end{array}\negthickspace\Biggl|\negthickspace\begin{array}{c}
1-m_{s_{\textrm{rd}}},\,1\\
m_{\textrm{rd}},\,0
\end{array}\negthickspace\Biggl|\negthickspace\begin{array}{c}
-1,-m_{s_{\textrm{sr}}},\,0\\
m_{\textrm{sr}}-1,\,0,\,-1
\end{array}\negthickspace\right]\Biggr)\label{eq:C2}
\end{eqnarray}

\selectlanguage{american}%
\centering{}\rule[0.5ex]{2.03\columnwidth}{0.8pt}\selectlanguage{english}%
\end{figure*}

\subsection{Asymptotic Performance Analysis}

In this subsection, we derive asymptotic expressions for the capacity
of the cooperative NOMA system. The asymptotic analysis is anchored
in obtaining expressions in higher SNR regimes to gain further insight
in the system performance. When the SNR $\gamma$ is sufficiently
large, such that $\ln\left(1+\gamma\right)\simeq\ln\left(\gamma\right),$
then the asymptotic capacity is given by the following lemma.
\begin{lem}
The asymptotic ergodic capacity of the cooperative NOMA system is
expressed as
\begin{equation}
\bar{C}_{\textrm{NOMA}}^{\textrm{asym}}=\bar{C}_{1,1}^{\textrm{asym}}-\bar{C}_{1,2}^{\textrm{asym}}+\bar{C}_{2}^{\textrm{asym}},\label{eq:NOMAasym-1}
\end{equation}
where $C_{1,1}^{\textrm{asym}}$ , $C_{1,2}^{\textrm{asym}}$ and
$C_{2}^{\textrm{asym}}$ are given by (\ref{eq:C11asymFinal}), (\ref{eq:C12asym})
and (\ref{eq:C2asym}), respectively.
\end{lem}
\begin{IEEEproof}
At high SNR, the instantaneous capacities in (\ref{eq:C11a}), (\ref{eq:C12a})
and (\ref{eq:C2main}) can be, respectively, reduced to
\begin{equation}
C_{1,1}^{\textrm{asym}}=\frac{1}{2}\,\textrm{log}_{2}\left(\textrm{min}\left\{ \bigl|h_{\textrm{sr}}\bigr|^{2},\,\bigl|h_{\textrm{sd}}\bigr|^{2}\right\} \bar{\gamma}\right),\label{eq:C11a-1}
\end{equation}
\begin{align}
C_{1,2}^{\textrm{asym}} & =\frac{1}{2}\,\textrm{log}_{2}\left(\textrm{min}\left\{ \bigl|h_{\textrm{sr}}\bigr|^{2},\,\bigl|h_{\textrm{sd}}\bigr|^{2}\right\} a_{2}\bar{\gamma}\right),\label{eq:C12a-1}
\end{align}
\begin{align}
C_{2}^{\textrm{asym}} & =\frac{1}{2}\,\textrm{log}_{2}\left(\textrm{min}\left\{ \bigl|h_{\textrm{sr}}\bigr|^{2}a_{2},\,\bigl|h_{\textrm{rd}}\bigr|^{2}\right\} \bar{\gamma}\right),\label{eq:C2main-1}
\end{align}
where $\bar{C}_{1,1}^{\textrm{asym}}=\stackrel[0]{\infty}{\int}f_{\mathcal{X}}\left(z\right)$
$\textrm{ln}\left(z\right)\textrm{d}z$, $\bar{C}_{1,2}^{\textrm{asym}}=\stackrel[0]{\infty}{\int}f_{\mathcal{Y}}\left(z\right)$
$\textrm{ln}\left(z\right)\textrm{d}z$ and $\bar{C}_{2}^{\textrm{asym}}$$=\stackrel[0]{\infty}{\int}f_{\mathcal{Z}}\left(z\right)$$\textrm{ln}\left(z\right)\textrm{d}z$. 

To compute $\bar{C}_{1,1}^{\textrm{asym}}$, we utilize (\ref{eq:PDFZ2})
to rewrite in the form presented in (\ref{eq:C11asym}), shown at
the top of this page 
\begin{figure*}[t]
\begin{eqnarray}
\bar{C}_{\textrm{1,1}}^{\textrm{asym}}=\underset{i\in\left\{ \textrm{sr},\textrm{sd}\right\} }{\sum}\frac{\Lambda_{i}^{m_{i}}}{B\left(m_{i},m_{s_{i}}\right)}\underset{J_{\textrm{1,1}}}{\underbrace{\stackrel[0]{\infty}{\int}\frac{z^{m_{i}-1}\textrm{ln}\left(z\right)}{\left(1+\Lambda_{i}z\right)^{m_{i}+m_{s_{i}}}}\textrm{d}z}}-\underset{k\in\left\{ \textrm{sr},\textrm{sd}\right\} }{\prod}\frac{\Lambda_{k}^{m_{k}}}{B\left(m_{k},m_{s_{k}}\right)}\qquad\hfill\qquad\qquad\qquad\qquad\qquad\nonumber \\
\qquad\hfill\qquad\qquad\qquad\:\times\underset{\underset{i\neq k}{i,l\in\left\{ \textrm{sr},\textrm{sd}\right\} }}{\sum}\frac{1}{m_{i}}\underset{J_{\textrm{1,2}}}{\underbrace{\stackrel[0]{\infty}{\int}\frac{z^{m_{i}+m_{l}-1}\textrm{ln}\left(z\right)}{\left(1+\Lambda_{l}z\right)^{m_{l}+m_{s_{l}}}}{}_{1}F_{2}\left(m_{i},m_{i}+m_{s_{i}},m_{i}+1;-\Lambda_{i}z\right)\textrm{d}z}}\label{eq:C11asym}
\end{eqnarray}

\selectlanguage{american}%
\centering{}\rule[0.5ex]{2.03\columnwidth}{0.8pt}\selectlanguage{english}%
\end{figure*}
. The integral $J_{\textrm{1,1}}$ in (\ref{eq:C11asym}) can be evaluated,
with the help of \cite[Eq. (4.293.14), (3.458)]{book2}, as\textcolor{black}{\allowdisplaybreaks
\begin{equation}
J_{\textrm{1,1}}=\frac{B\left(m_{i},m_{s_{i}}\right)}{\Lambda_{i}^{m_{i}}}\left(\psi^{\left(0\right)}\left(m_{i}\right)-\psi^{\left(0\right)}\left(m_{s_{i}}\right)-\textrm{ln}\left(\Lambda_{i}\right)\right).\label{eq:J11}
\end{equation}
}

Now, to solve the integral $J_{\textrm{1,2}}$ in (\ref{eq:C11asym}),
we first expand the Gauss hypergeometric function in terms of the
series representation \textcolor{black}{\cite[Eq. (9.14.1)]{book2}}.
Then exchanging the integral and summation order along with some mathematical
manipulations, the integral $J_{\textrm{1,2}}$ can be rewritten as
\textcolor{black}{\allowdisplaybreaks}
\begin{align}
J_{\textrm{1,2}}= & \stackrel[n=0]{\infty}{\sum}\frac{\left(m_{i}+m_{s_{i}}\right)_{n}\Lambda_{i}^{n}}{\left(m_{i}+1\right)_{n}}\nonumber \\
 & \times\underset{J_{\textrm{1,3}}}{\underbrace{\stackrel[0]{\infty}{\int}\frac{z^{m_{i}+m_{l}+n-1}\textrm{ln}\left(z\right)}{\left(1+\Lambda_{l}z\right)^{m_{l}+m_{s_{l}}}\left(1+\Lambda_{i}z\right)^{m_{i}+m_{s_{i}}+n}}\textrm{d}z}}.\label{eq:J12}
\end{align}

Note that the transformation in (\ref{eq:Transformation-Gauss}) was
utilized to arrive at (\ref{eq:J12}). To the authors' best knowledge,
the integral $J_{\textrm{1,3}}$ can not be solved in closed-form
in its current form. However, assuming that $m_{s_{i}}-1=Cm_{i}$,
where $C\in\mathbb{N}$, we get 
\begin{align}
J_{\textrm{1,3}}= & \stackrel[0]{\infty}{\int}\frac{z^{m_{i}+m_{l}+n-1}\textrm{ln}\left(z\right)}{\left(1+\frac{z}{C\bar{\gamma}}\right)^{m_{l}+m_{s_{l}}+m_{i}+m_{s_{i}}+n}}\textrm{d}z,\nonumber \\
\stackrel{\left(a\right)}{=} & \left(C\bar{\gamma}\right)^{m_{l}+m_{i}+n}B\left(m_{s_{l}}+m_{s_{i}},m_{i}+m_{l}+n\right)\nonumber \\
 & \left(\psi^{\left(0\right)}\left(m_{i}+m_{l}+n\right)-\psi^{\left(0\right)}\left(m_{s_{i}}+m_{s_{l}}\right)+\textrm{ln}\left(C\bar{\gamma}\right)\right).\label{eq:J13}
\end{align}

Now, substituting (\ref{eq:J13}) into (\ref{eq:J12}), then (\ref{eq:J12})
and (\ref{eq:J11}) into (\ref{eq:C11asym}), along with some basic
algebra, yields
\begin{eqnarray}
\bar{C}_{1,1}^{\textrm{asym}}=\underset{i\in\left\{ \textrm{sr},\textrm{sd}\right\} }{\sum}\left(\psi^{\left(0\right)}\left(m_{i}\right)-\psi^{\left(0\right)}\left(m_{s_{i}}\right)-\textrm{ln}\left(\Lambda_{i}\right)\right)\quad\quad\nonumber \\
\negthickspace\negthickspace\negthickspace\negthickspace\negthickspace\negthickspace\negthickspace\negthickspace\negthickspace\negthickspace\negthickspace\negthickspace\negthickspace\negthickspace\negthickspace\negthickspace\negthickspace\negthickspace\negthickspace\negthickspace\negthickspace\negthickspace\negthickspace\negthickspace\negthickspace\negthickspace\negthickspace\negthickspace-\underset{\underset{i\neq l}{i,l\in\left\{ \textrm{sr},\textrm{sd}\right\} }}{\sum}\stackrel[n=0]{\infty}{\sum}\frac{B\left(m_{i}+m_{l}+n,m_{s_{i}}+m_{s_{l}}\right)}{B\left(m_{i},m_{s_{l}}\right)B\left(m_{i}+n,m_{s_{i}}\right)\left(m_{l}+n\right)}\nonumber \\
\times\left(\psi^{\left(0\right)}\left(m_{i}+m_{l}+n\right)-\psi^{\left(0\right)}\left(m_{s_{i}}+m_{s_{l}}\right)-\textrm{ln}\left(\bar{\gamma}C\right)\right)\label{eq:C11asymFinal}
\end{eqnarray}
Following the same procedure used to derive (\ref{eq:C11asymFinal}),
it is straightforward to show that $\bar{C}_{1,2}^{\textrm{asym}}$
can be given as in (\ref{eq:C12asym}), shown at the top of this page.
\begin{figure*}[t]
\begin{eqnarray}
\bar{C}_{1,2}^{\textrm{asym}}=\underset{i\in\left\{ \textrm{sr},\textrm{sd}\right\} }{\sum}\left(\psi^{\left(0\right)}\left(m_{i}\right)-\psi^{\left(0\right)}\left(m_{s_{i}}\right)-\textrm{ln}\left(\frac{\Lambda_{i}}{a_{2}}\right)\right)-\underset{\underset{i\neq l}{i,l\in\left\{ \textrm{sr},\textrm{sd}\right\} }}{\sum}\stackrel[n=0]{\infty}{\sum}\frac{B\left(m_{i}+m_{l}+n,m_{s_{i}}+m_{s_{l}}\right)}{B\left(m_{i},m_{s_{l}}\right)B\left(m_{i}+n,m_{s_{i}}\right)}\nonumber \\
\qquad\qquad\:\times\frac{1}{\left(m_{l}+n\right)}\left(\psi^{\left(0\right)}\left(m_{i}+m_{l}+n\right)-\psi^{\left(0\right)}\left(m_{s_{i}}+m_{s_{l}}\right)-\textrm{ln}\left(\bar{\gamma}Ca_{2}\right)\right)\hspace*{1em}\qquad\hfill\label{eq:C12asym}
\end{eqnarray}

\selectlanguage{american}%
\centering{}\rule[0.5ex]{2.03\columnwidth}{0.8pt}\selectlanguage{english}%
\end{figure*}
. Similarly, we can show that $C_{2}^{\textrm{asym}}$ is given by
\begin{multline}
\bar{C}_{2}^{\textrm{asym}}=\underset{i\in\left\{ \textrm{sr},\textrm{sd}\right\} }{\sum}\left(\psi^{\left(0\right)}\left(m_{i}\right)-\psi^{\left(0\right)}\left(m_{s_{i}}\right)-\textrm{ln}\left(\frac{\Lambda_{i}}{a_{2}}\right)\right)\\
-\underset{\underset{i\neq l}{i,l\in\left\{ \textrm{sr},\textrm{sd}\right\} }}{\sum}\stackrel[n=0]{\infty}{\sum}\frac{B\left(m_{i}+m_{l}+n,m_{s_{i}}+m_{s_{l}}\right)}{B\left(m_{i},m_{s_{l}}\right)B\left(m_{i}+n,m_{s_{i}}\right)\left(m_{l}+n\right)}\\
\times\left(\psi^{\left(0\right)}\left(m_{i}+m_{l}+n\right)-\psi^{\left(0\right)}\left(m_{s_{i}}+m_{s_{l}}\right)-\textrm{ln}\left(\bar{\gamma}Ca_{2}\right)\right).\label{eq:C2asym}
\end{multline}
 This concludes the proof.
\end{IEEEproof}

\section{\label{sec:Cooperative-OMA} Cooperative OMA over $\mathcal{F}$
Fading Channels }

In this section, we analyze the exact and asymptotic ergodic sum rate
performance of the cooperative OMA scheme with DF relaying over the
Fisher-Snedecor $\mathcal{F}$ composite fading channel. 

\subsection{Exact Performance Analysis }

The instantaneous capacity of this system is given by \cite{InforTh_DF79,DF_WC09,NOMAmain}

\vspace*{2mm} 

\noindent 
\begin{equation}
C_{\textrm{OMA}}=\frac{1}{2}\,\textrm{log}_{2}\left(1+\textrm{min}\left\{ \bigl|h_{\textrm{sr}}\bigr|^{2},\,\bigl|h_{\textrm{sd}}\bigr|^{2}+\bigl|h_{\textrm{rd}}\bigr|^{2}\right\} \bar{\gamma}\right).\label{eq:Coma}
\end{equation}

Let $\mathcal{W}=\textrm{min}\left\{ \mathcal{W}_{\textrm{sr}},\,\mathcal{W}_{\textrm{sd}}+\mathcal{W}_{\textrm{rd}}\right\} $,
where $\mathcal{W}_{i}=\bigl|h_{i}\bigr|^{2}\bar{\gamma}$ and $i\in\left\{ \textrm{sr},\,\textrm{sd},\,\textrm{rd}\right\} $,
then the ergodic capacity, $\bar{C}_{\textrm{OMA}}$, can be calculated
as 

\begin{equation}
\bar{C}_{\textrm{OMA}}=\stackrel[0]{\infty}{\int}\textrm{log}_{2}\left(1+u\right)f_{\mathcal{W}}\left(u\right)\textrm{d}u=\frac{1}{2\textrm{ln}\left(2\right)}\stackrel[0]{\infty}{\int}\frac{F_{\mathcal{W}}\left(u\right)}{1+u}\textrm{d}u,\label{eq:Coma2}
\end{equation}

\noindent where $F_{\mathcal{W}}\left(\cdot\right)$ is the CDF of
the RV $\mathcal{W}$, which is given by 

\begin{equation}
F_{\mathcal{W}}\left(u\right)=1-\bar{F}_{\mathcal{W}_{\textrm{sr}}}\left(u\right)\,\bar{F}_{\mathcal{W}_{\textrm{sd}}+\mathcal{W}_{\textrm{rd}}}\left(u\right),\label{eq:CDFW}
\end{equation}

\noindent with $\bar{F}_{\mathcal{W}_{\textrm{sr}}}\left(u\right)$
representing the CCDF of $\mathcal{W}_{\textrm{sr}}$ and $\bar{F}_{\mathcal{W}_{\textrm{sd}}+\mathcal{W}_{\textrm{rd}}}\left(u\right)$
denotes the CCDF of the sum of two Fisher-Snedecor $\mathcal{F}$
variates, $\mathcal{W}_{\textrm{sd}}+\mathcal{W}_{\textrm{rd}}$.

Since the RV $\mathcal{W}_{\textrm{sr}}$ follows the Fisher-Snedecor
$\mathcal{F}$ distribution, its CDF $F_{\mathcal{W}_{\textrm{sr}}}\left(\cdot\right)$
can be obtained directly from (\ref{eq:CDFXY}), with the appropriate
notation changes, as

\vspace*{2.5mm} 

\noindent 
\begin{equation}
F_{\mathcal{W}_{\textrm{sr}}}\left(z\right)=\frac{1}{\Gamma\left(m_{\textrm{sr}}\right)\Gamma\left(m_{s_{\textrm{sr}}}\right)}\,G_{2,2}^{1,2}\left[\Lambda_{\textrm{sr}}z\Biggl|\negthickspace\begin{array}{c}
1-m_{s_{\textrm{sr}}},\,1\\
m_{\textrm{sr}},\,0
\end{array}\negthickspace\right],\label{eq:CDFX1-1}
\end{equation}

\noindent which, after invoking \cite[Eq. (17)]{Adamchik90}, can
be expressed in terms of the Gauss hypergeometric function as

\begin{equation}
F_{\mathcal{W}_{\textrm{sr}}}\left(z\right)=\Psi z^{m_{\textrm{sr}}}{}_{2}F_{1}\left(m_{\textrm{sr}},m_{\textrm{sr}}+m_{s_{\textrm{sr}}};m_{\textrm{sr}}+1;-\Lambda_{\textrm{sr}}z\right),\label{eq:FW1}
\end{equation}

\noindent where $\Psi=\frac{\Lambda_{\textrm{sr}}^{m_{\textrm{sr}}}}{\textrm{Beta}\left[m_{\textrm{sr}},m_{s_{\textrm{sr}}}\right]m_{\textrm{sr}}}$.

As for the CDF of $\mathcal{W}_{\textrm{sd}}+\mathcal{W}_{\textrm{rd}}$,
fortunately, the characterization of the sum of Fisher-Snedecor $\mathcal{F}$
variates has, very recently, been studied in \cite{badarneh2019sum2}.
More specifically, using \cite[Eq. (8)]{badarneh2019sum2}, we can
express $F_{\mathcal{W}_{\textrm{sd}}+\mathcal{W}_{\textrm{rd}}}\left(\cdot\right)$
as follows 

\begin{align}
F_{\mathcal{W}_{\textrm{sd}}+\mathcal{W}_{\textrm{rd}}}\left(z\right) & =\Upsilon z^{2m_{\textrm{rd}}}\nonumber \\
\times & _{2}F_{1}\left(m_{\textrm{rd}}+m_{s_{\textrm{rd}}},2m_{\textrm{rd}};2m_{\textrm{rd}}+1;-\Lambda_{\textrm{rd}}z\right),\label{eq:FW2}
\end{align}

\noindent where $\Upsilon=\frac{\Gamma\left[\varpi_{\textrm{rd}}\right]^{2}\Lambda_{\textrm{rd}}^{2m_{\textrm{rd}}}}{\Gamma\left[m_{s_{\textrm{rd}}}\right]^{2}\Gamma\left[2m_{\textrm{rd}}+1\right]}$
and $\varpi_{i}=m_{i}+m_{s_{i}}$. It should be mentioned that (\ref{eq:FW2})
is based on the fact that $\mathcal{W}_{\textrm{sd}}$ and $\mathcal{W}_{\textrm{rd}}$
are independent and identically distributed Fisher-Snedecor $\mathcal{F}$
variates. 

Now, substituting (\ref{eq:FW1}) and (\ref{eq:FW2}) into (\ref{eq:CDFW})
and then into (\ref{eq:Coma2}), with some algebraic manipulations,
we can express $\bar{C}_{\textrm{OMA}}$ as 

\begin{align}
\bar{C}_{\textrm{OMA}} & =\frac{1}{2\textrm{ln}\left(2\right)}\left(\mathcal{J}_{1}-\Upsilon\mathcal{J}_{2}-\Psi\mathcal{J}_{3}+\Psi\Upsilon\mathcal{J}_{4}\right),\label{eq:Coma-1}
\end{align}

\noindent where $\mathcal{J}_{1},\,\mathcal{J}_{2},\,\mathcal{J}_{3}\,\textrm{and}\,\mathcal{J}_{4}$
are integrals given respectively as

\begin{align}
\mathcal{J}_{1} & =\stackrel[0]{\upsilon}{\int}\frac{1}{1+z}\textrm{d}z=\textrm{ln}\left(1+\upsilon\right),\label{eq:I1_final}
\end{align}

\begin{align}
\mathcal{J}_{2} & =\stackrel[0]{\upsilon}{\int}\frac{z^{2m_{\textrm{rd}}}}{1+z}{}_{2}F_{1}\left(\varpi_{\textrm{rd}},2m_{\textrm{rd}};2m_{\textrm{rd}}+1;-\Lambda_{\textrm{rd}}z\right)\textrm{d}z,\label{eq:I2-1}
\end{align}

\begin{align}
\mathcal{J}_{3} & =\stackrel[0]{\upsilon}{\int}\frac{z^{m_{\textrm{sr}}}}{1+z}{}_{2}F_{1}\left(m_{\textrm{sr}},\varpi_{\textrm{sr}};m_{\textrm{sr}}+1;-\Lambda_{\textrm{sr}}z\right)\textrm{d}z,
\end{align}

\begin{align}
\mathcal{J}_{4}= & \stackrel[0]{\upsilon}{\int}\frac{z^{m_{\textrm{sr}}+2m_{\textrm{rd}}}}{1+z}{}_{2}F_{1}\left(\varpi_{\textrm{rd}},2m_{\textrm{rd}};2m_{\textrm{rd}}+1;-\Lambda_{\textrm{rd}}z\right)\nonumber \\
 & \qquad\times{}_{2}F_{1}\left(m_{\textrm{sr}},\varpi_{\textrm{sr}};m_{\textrm{sr}}+1;-\Lambda_{\textrm{sr}}z\right)\textrm{d}z,\label{eq:I4}
\end{align}

\noindent where $\upsilon=1/\Lambda_{\textrm{rd}}$. 

We now evaluate the integral $\mathcal{J}_{2}$. It should be noted
that the Gauss hypergeometric function in (\ref{eq:I2-1}) converges
onl\textcolor{black}{y when $\bigl|\Lambda_{\textrm{rd}}<1\bigr|$.
The}refore, in order to overcome this restriction, we use the following
transformation \cite[Eq. (7.2.4.36)]{bookV3} 

\begin{align}
_{2}F_{1}\left(a,b;c;x\right) & =\left(1-x\right)^{-b}\,_{2}F_{1}\left(c-a;b;c;\frac{x}{x-1}\right).\label{eq:Transformation-Gauss}
\end{align}

By using (\ref{eq:Transformation-Gauss}) an\textcolor{black}{d expanding
the Gauss hypergeometric function in terms of the series representation
\cite[Eq. (9.14.1)]{book2}, along with some} simple mathematical
manipulations, we can rewrite $\mathcal{J}_{2}$ as 

\begin{align}
\mathcal{J}_{2} & =\stackrel[n=0]{\infty}{\sum}\frac{\left(m_{\textrm{rd}}-m_{s_{\textrm{rd}}}+1\right)_{n}\left(2m_{\textrm{rd}}\right)_{n}}{n!\left(2m_{\textrm{rd}}+1\right)_{n}\Lambda_{\textrm{rd}}^{-n}}\qquad\qquad\nonumber \\
 & \qquad\hfill\qquad\times\underset{\mathcal{J}_{2,1}}{\underbrace{\stackrel[0]{\upsilon}{\int}\frac{z^{2m_{\textrm{rd}}+n}}{1+z}\left(\frac{1}{\Lambda_{\textrm{rd}}z+1}\right)^{2m_{\textrm{rd}}+n}\textrm{d}z}},\label{eq:I2-2}
\end{align}

\noindent where $\left(x\right)_{k}=\Gamma\left(x+k\right)/\Gamma\left(x\right)$
denotes the Pochhammer symbol defined as

\begin{equation}
\left(b\right)_{n}=\biggl\{\begin{array}{cc}
b\left(b+1\right)\ldots\left(b+n-1\right), & n\in N\\
1.\hfill\hfill & n\in0
\end{array}
\end{equation}

The integral $\mathcal{J}_{2,1}$ in (\ref{eq:I2-2}) has the form
$\int y^{r}$ $\left(1+ay\right)^{p}$$\left(1+by\right)^{q}$ which
can be evaluated straightforwardly in closed-form in terms of the
Appell hypergeometric function. More specifically, using 

\begin{align}
\stackrel[0]{c}{\int}y^{\alpha} & \left(\frac{1}{1+ay}\right)^{\beta_{1}}\left(\frac{1}{1+by}\right)^{\beta_{2}}\textrm{d}z=\frac{c^{\alpha+1}}{\alpha+1}\nonumber \\
 & \times F_{1}\left(\alpha+1;\beta_{1},\beta_{2};\alpha+2;-ac,-bc\right),\label{eq:F1}
\end{align}

\noindent we can evaluate $\mathcal{J}_{2,1}$ as 

\begin{align}
\mathcal{J}_{2,1}= & \frac{\Lambda_{\textrm{rd}}^{-\xi}}{\xi}F_{1}\left(\xi;2m_{m_{\textrm{rd}}}+n,1;\psi+1;-1,-\frac{1}{\Lambda_{\textrm{rd}}}\right),\label{eq:Appell}
\end{align}

\noindent where $\xi=2m_{\textrm{rd}}+n+1.$ Note that the Appell
function in (\ref{eq:Appell}) converges only when $\bigl|\frac{1}{\Lambda_{\textrm{rd}}}<1\bigr|$.
To overcome this restriction, we use the transformation \cite[Eq. (7.2.4.36)]{bookV3} 

\begin{align}
F_{1}\left(a;b_{1},b_{2};c;x_{1},x_{2}\right) & =\underset{i\in\left(1,2\right)}{\prod}\left(1-x_{i}\right)^{-b_{i}}\nonumber \\
\times F_{1}\Biggl(c-a; & b_{1},b_{2};c;\frac{x_{1}}{x_{1}-1},\frac{x_{2}}{x_{2}-1}\Biggr).\label{eq:Transformation}
\end{align}

Using (\ref{eq:Transformation}) and (\ref{eq:Appell}), along with
some algebraic manipulations, we can express $\mathcal{J}_{2}$ in
closed-form as follows 

\begin{align}
\mathcal{J}_{2}= & \stackrel[n=0]{\infty}{\sum}\frac{\left(m_{\textrm{rd}}-m_{s_{\textrm{rd}}}+1\right)_{n}\left(2m_{\textrm{rd}}\right)_{n}\Lambda_{\textrm{rd}}^{2m_{\textrm{rd}}}}{n!\left(2m_{\textrm{rd}}+1\right)_{n}\left(\Lambda_{\textrm{rd}}+1\right)2^{\left(2m_{\textrm{rd}}+n\right)}\xi}\nonumber \\
 & \qquad\times F_{1}\left(1;2m_{m_{\textrm{rd}}}+n,1;\xi+1;\frac{1}{2},\frac{1}{\Lambda_{\textrm{rd}}+1}\right).\label{eq:I2final}
\end{align}

Following the same procedure used to obtain (\ref{eq:I2final}), it
is straightforward to evaluate the integral $\mathcal{J}_{3}$. Using
the transformation (\ref{eq:Transformation-Gauss}) and the series
representation of $_{2}F_{1}$ \cite[Eq. (9.14.1)]{book2}, we can
rewrite $\mathcal{J}_{3}$ as

\begin{align}
\mathcal{J}_{3} & =\stackrel[l=0]{\infty}{\sum}\frac{\left(\varpi_{\textrm{sr}}\right)_{l}\Lambda_{\textrm{sr}}^{l}}{\left(m_{\textrm{sr}}+1\right)_{l}}\underset{\mathcal{J}_{3,1}}{\underbrace{\stackrel[0]{\upsilon}{\int}\frac{z^{-m_{s_{\textrm{sr}}}}}{1+z}\left(\frac{z}{\Lambda_{\textrm{sr}}z+1}\right)^{\varpi_{\textrm{sr}}+l}\textrm{d}z}}.\label{eq:I3-1}
\end{align}

With the aid of (\ref{eq:F1}), we can evaluate the integral in (\ref{eq:I3-1})
as 

\begin{align}
\mathcal{J}_{3,1}= & \frac{\Lambda_{\textrm{rd}}^{-\phi}}{\phi}\,F_{1}\left(\phi;\varpi_{\textrm{sr}}+l,1;\phi+1;-\frac{\Lambda_{\textrm{sr}}}{\Lambda_{\textrm{rd}}},-\frac{1}{\Lambda_{\textrm{rd}}}\right),\label{eq:Appell-1}
\end{align}

\noindent where $\phi=m_{\textrm{sr}}+l+1.$ 

Substituting (\ref{eq:Appell-1}) into (\ref{eq:I3-1}), along with
some mathematical manipulations, we can express the integral $\mathcal{J}_{3}$
in closed-form as given in (\ref{eq:I3final}), shown at the top of
the next page. Note that we used $\left(x\right)_{n}=\frac{\Gamma\left(x+n\right)}{\Gamma\left(x\right)}\,\textrm{and}\,\Gamma\left(x+1\right)=x!$
to arrive at (\ref{eq:I3final}).

\begin{figure*}[t]
\begin{align}
\mathcal{J}_{3}=\stackrel[l=0]{\infty}{\sum}\frac{\Gamma\left(\varpi_{\textrm{sr}}+l\right)\Gamma\left(m_{\textrm{sr}}+1\right)\Lambda_{\textrm{sr}}^{l}}{\Gamma\left(\phi+1\right)\Gamma\left(\varpi_{\textrm{sr}}\right)\Lambda_{\textrm{rd}}^{\phi}}\,F_{1}\left(m_{\textrm{sr}}+l+1;\varpi_{\textrm{sr}}+l,1;m_{\textrm{sr}}+l+2;-\frac{\Lambda_{\textrm{sr}}}{\Lambda_{\textrm{rd}}},-\frac{1}{\Lambda_{\textrm{rd}}}\right)\label{eq:I3final}
\end{align}

\selectlanguage{american}%
\centering{}\rule[0.5ex]{2.03\columnwidth}{0.8pt}\selectlanguage{english}%
\end{figure*}

Furthermore, to solve the integral $\mathcal{J}_{4}$, we first replace
the two Gauss hypergeometric functions in (\ref{eq:I4}) with their
series representations, \cite[Eq. (9.14.1)]{book2}, after applying
 the transformation in (\ref{eq:Transformation-Gauss}). This yields
(\ref{eq:I4-1}), shown at the top of the next page. Unfortunately,
it is very difficult to evaluate the integral $\mathcal{J}_{\textrm{4,1}}$
in (\ref{eq:I4-1}) and, to the best of our knowledge, it can not
be expressed in closed-form in its current form. However, letting
$\left(m_{s_{i}}-1\right)=Cm_{i}$, where $i\in\left\{ \textrm{sr},\textrm{rd}\right\} $
and $C\in R$, we can simplify this integral to 

\begin{figure*}[t]
\begin{align}
\mathcal{J}_{4}=\stackrel[k=0]{\infty}{\sum}\stackrel[m=0]{\infty}{\sum}\frac{\left(m_{\textrm{rd}}-m_{s_{\textrm{rd}}}+1\right)_{k}\left(2m_{\textrm{rd}}\right)_{k}}{\left(2m_{\textrm{rd}}+1\right)_{k}\Lambda_{\textrm{rd}}^{-n}k!}\frac{\left(\varpi_{\textrm{sr}}\right)_{m}\Lambda_{\textrm{sr}}^{m}}{\left(m_{\textrm{sr}}+1\right)_{m}} & \underset{\mathcal{J}_{\textrm{4,1}}}{\underbrace{\stackrel[0]{\upsilon}{\int}\frac{z^{m_{\textrm{sr}}+2m_{\textrm{rd}}+k+m}}{1+z}\left(\frac{1}{\Lambda_{\textrm{rd}}z+1}\right)^{2m_{\textrm{rd}}+k}\left(\frac{1}{\Lambda_{\textrm{sr}}z+1}\right)^{\varpi_{\textrm{sr}}+m}\textrm{d}z}}\label{eq:I4-1}
\end{align}

\selectlanguage{american}%
\centering{}\rule[0.5ex]{2.03\columnwidth}{0.8pt}\selectlanguage{english}%
\end{figure*}

\begin{equation}
\mathcal{J}_{\textrm{4,1}}=\frac{\Lambda_{\textrm{rd}}^{-\left(2m_{\textrm{rd}}+n\right)}}{\Lambda_{\textrm{sr}}^{\varpi_{\textrm{sr}}+m}}\stackrel[0]{\upsilon}{\int}\frac{z^{\kappa}}{1+z}\left(\frac{1}{z+C\bar{\gamma}}\right)^{m_{s_{\textrm{sr}}}+\kappa}\textrm{d}z,
\end{equation}

\noindent where $\kappa=m_{\textrm{sr}}+2m_{\textrm{rd}}+k+m$. 

Now, with the aid of (\ref{eq:F1}), $\mathcal{J}_{\textrm{4,1}}$
can be evaluated as

\begin{equation}
\mathcal{J}_{\textrm{4,1}}=\frac{\left(\bar{\gamma}C\right)^{\kappa+1}}{\kappa+1}\,F_{1}\left(\kappa+1;m_{s_{\textrm{sr}}}+\kappa,1;\kappa+2;-1,-\bar{\gamma}C\right).\label{eq:I42}
\end{equation}

Now, substituting (\ref{eq:I42}) into (\ref{eq:I4-1}), along with
some rearrangements, yields (\ref{eq:I4-3}), shown at the top of
the next page. Finally, substituting (\ref{eq:I1_final}), (\ref{eq:I2final}),
(\ref{eq:I3final}) and (\ref{eq:I4-3}) into (\ref{eq:Coma-1}),
we obtain the ergodic capacity of the cooperative OMA system over
the Fisher-Snedecor $\mathcal{F}$ composite fading channel.
\begin{figure*}[t]
\begin{align}
\mathcal{J}_{4}=\stackrel[k=0]{\infty}{\sum}\stackrel[m=0]{\infty}{\sum}\frac{\left(\varpi_{\textrm{sr}}\right)_{m}\left(m_{\textrm{rd}}-m_{s_{\textrm{rd}}}+1\right)_{k}2m_{\textrm{rd}}}{\left(m_{\textrm{sr}}+1\right)_{m}\left(2m_{\textrm{rd}}+n\right)\left(\kappa+1\right)\left(\bar{\gamma}C\right)^{-\left(m_{\textrm{sr}}+2m_{\textrm{rd}}+1\right)}k!} & F_{1}\left(\kappa+1;m_{s_{\textrm{sr}}}+\kappa,1;\kappa+2;-1,-\bar{\gamma}C\right)\label{eq:I4-3}
\end{align}

\selectlanguage{american}%
\centering{}\rule[0.5ex]{2.03\columnwidth}{0.8pt}\selectlanguage{english}%
\end{figure*}

\subsection{Asymptotic Performance Analysis}

In this section, we analysis the asymptotic capacity performance of
the cooperative OMA system. At high SNR, the instantaneous capacity
in (\ref{eq:Coma}) can be simplified to 

\vspace*{2mm} 

\noindent 
\begin{equation}
C_{\textrm{OMA}}^{\textrm{asym}}=\frac{1}{2}\,\textrm{log}_{2}\left(\textrm{min}\left\{ \bigl|h_{\textrm{sr}}\bigr|^{2},\,\bigl|h_{\textrm{sd}}\bigr|^{2}+\bigl|h_{\textrm{rd}}\bigr|^{2}\right\} \bar{\gamma}\right).\label{eq:Comaasym}
\end{equation}

Similar to the analysis in Sec. \ref{sec:Cooperative-OMA}, while
assuming high SNR regime, it is easy to show that the asymptotic ergodic
capacity for the OMA can be given by 

\begin{align}
\bar{C}_{\textrm{OMA}}^{\textrm{asym}} & =\frac{1}{2\textrm{ln}\left(2\right)}\left(\mathcal{J}_{1}^{\textrm{asym}}-\Upsilon\mathcal{J}_{2}^{\textrm{asym}}-\Psi\mathcal{J}_{3}^{\textrm{asym}}+\Psi\Upsilon\mathcal{J}_{4}^{\textrm{asym}}\right),\label{eq:OMAasym}
\end{align}

\noindent where $\mathcal{J}_{1}^{\textrm{asym}},\mathcal{J}_{2}^{\textrm{asym}},\mathcal{J}_{3}^{\textrm{asym}}\,\textrm{and}\,\mathcal{J}_{4}^{\textrm{asym}}$
are given by $\mathcal{J}_{1}^{\textrm{asym}}=\stackrel[1]{\upsilon}{\int}\frac{1}{z}$$\textrm{d}z$$=\textrm{ln}\left(\upsilon\right)$, 

\begin{align}
\mathcal{J}_{2}^{\textrm{asym}} & =\stackrel[0]{\upsilon}{\int}z^{2m_{\textrm{rd}}-1}{}_{2}F_{1}\left(\varpi_{\textrm{rd}},2m_{\textrm{rd}};2m_{\textrm{rd}}+1;-\Lambda_{\textrm{rd}}z\right)\textrm{d}z,\label{eq:I2Asym}
\end{align}

\begin{align}
\mathcal{J}_{3}^{\textrm{asym}} & =\stackrel[0]{\upsilon}{\int}z^{m_{\textrm{sr}}-1}{}_{2}F_{1}\left(m_{\textrm{sr}},\varpi_{\textrm{sr}};m_{\textrm{sr}}+1;-\Lambda_{\textrm{sr}}z\right)\textrm{d}z,\label{eq:I3Asym-2}
\end{align}

\begin{align}
\mathcal{J}_{4}^{\textrm{asym}}= & \stackrel[0]{\upsilon}{\int}z^{m_{\textrm{sr}}+2m_{\textrm{rd}}-1}{}_{2}F_{1}\left(\varpi_{\textrm{rd}},2m_{\textrm{rd}};2m_{\textrm{rd}}+1;-\Lambda_{\textrm{rd}}z\right)\nonumber \\
 & \qquad\qquad\times{}_{2}F_{1}\left(m_{\textrm{sr}},\varpi_{\textrm{sr}};m_{\textrm{sr}}+1;-\Lambda_{\textrm{sr}}z\right)\textrm{d}z.\label{eq:I4Asym}
\end{align}

The integral $\mathcal{J}_{2}^{\textrm{asym}}$ in (\ref{eq:I2Asym}),
after using the transformation in (\ref{eq:Transformation-Gauss}),
can be solved as 
\begin{align}
\mathcal{J}_{2}^{\textrm{asym}}= & \frac{1}{2m_{\textrm{rd}}\Lambda_{\textrm{rd}}^{2m_{\textrm{rd}}}}\,{}_{3}F_{1}\left(2m_{\textrm{rd}},2m_{\textrm{rd}},\varpi_{\textrm{rd}};A_{\textrm{rd}},A_{\textrm{rd}},-1\right),\label{eq:I2Asym-1}
\end{align}

\noindent where $A_{\textrm{rd}}=2m_{\textrm{rd}}+1$ and $_{p}F_{q}\left(\cdot,\cdot,\cdot;\cdot,\cdot,\cdot\right)$
is the generalized hypergeometric function \cite[Eq. (9.14.1)]{book2}.
Similarly, it does not pose any difficulty to show that 

\begin{align}
\mathcal{J}_{3}^{\textrm{asym}}= & \frac{1}{m_{\textrm{sr}}\Lambda_{\textrm{rd}}^{m_{\textrm{sr}}}}\,{}_{3}F_{2}\left(m_{\textrm{sr}},m_{\textrm{sr}},\varpi_{\textrm{sr}};A_{\textrm{sr}},A_{\textrm{sr}},-\frac{\Lambda_{\textrm{sr}}}{\Lambda_{\textrm{rd}}}\right),\label{eq:I3Asym-1}
\end{align}

\noindent where $A_{\textrm{sr}}=2m_{\textrm{sr}}+1$. 

Now, to solve the integral $\mathcal{J}_{4}^{\textrm{asym}}$, we
first apply the transformation in (\ref{eq:Transformation-Gauss})
and then replace the two hypergeometric functions with their series
representation \cite[Eq. (9.14.1)]{book2}. Along with some basic
algebra and reordering of integration and summations, we can rewrite
(\ref{eq:I4Asym}) as

\begin{align}
\mathcal{J}_{4}^{\textrm{asym}}=\stackrel[k=0]{\infty}{\sum}\stackrel[m=0]{\infty}{\sum}\frac{\left(m_{\textrm{rd}}-m_{s_{\textrm{rd}}}+1\right)_{k}\left(2m_{\textrm{rd}}\right)_{k}}{\left(2m_{\textrm{rd}}+1\right)_{k}\Lambda_{\textrm{rd}}^{-n}k!}\frac{\left(\varpi_{\textrm{sr}}\right)_{m}\Lambda_{\textrm{sr}}^{m}}{\left(m_{\textrm{sr}}+1\right)_{m}}\nonumber \\
\times\underset{\mathcal{J}_{4,1}^{\textrm{asym}}}{\underbrace{\stackrel[0]{\upsilon}{\int}z^{\kappa-1}\left(\frac{1}{\Lambda_{\textrm{rd}}z+1}\right)^{2m_{\textrm{rd}}+k}\left(\frac{1}{\Lambda_{\textrm{sr}}z+1}\right)^{\varpi_{\textrm{sr}}+m}\textrm{d}z}}.\label{eq:I4Asym-1}
\end{align}

With the help of (\ref{eq:F1}), we can straightforwardly evaluate
the integral $\mathcal{J}_{4,1}^{\textrm{asym}}$ in closed-form.
At this end, we can now write $\mathcal{J}_{4}^{\textrm{asym}}$ as
in (\ref{eq:I4Asym3}), given at the top of the next page.
\begin{figure*}[t]
\begin{align}
\mathcal{J}_{4}^{\textrm{asym}}=\stackrel[k=0]{\infty}{\sum}\stackrel[m=0]{\infty}{\sum}\frac{\left(m_{\textrm{rd}}-m_{s_{\textrm{rd}}}+1\right)_{k}\left(2m_{\textrm{rd}}\right)_{k}\left(\varpi_{\textrm{sr}}\right)_{m}}{\left(2m_{\textrm{rd}}+1\right)_{k}\left(m_{\textrm{sr}}+1\right)_{m}k!}\frac{\Lambda_{\textrm{sr}}^{m}}{\kappa\Lambda_{\textrm{rd}}^{\kappa-k}} & \,F_{1}\left(\kappa;\varpi_{\textrm{sr}}+m,2m_{\textrm{rd}}+k;\kappa+1;-\frac{\Lambda_{\textrm{sr}}}{\Lambda_{\textrm{rd}}},-1\right)\label{eq:I4Asym3}
\end{align}

\selectlanguage{american}%
\centering{}\rule[0.5ex]{2.03\columnwidth}{0.8pt}\selectlanguage{english}%
\end{figure*}

Finally, substituting $\mathcal{J}_{1}^{\textrm{asym}}$, $\mathcal{J}_{2}^{\textrm{asym}}$,
$\mathcal{J}_{3}^{\textrm{asym}}$ and $\mathcal{J}_{4}^{\textrm{asym}}$
into (\ref{eq:OMAasym}) yields a closed-form expression for the asymptotic
ergodic capacity of the cooperative OMA system over Fisher-Snedecor
$\mathcal{F}$ composite fading channels.

\section{\textcolor{black}{\label{sec:Numerical-Examples}Results and Discussions}}

In this section, we present some numerical examples of the analytical
expressions derived above along with Monte-Carlo simulations. All
evaluations herein, unless specified otherwise, are based on the following
system parameters: $m_{\textrm{sr}}=m_{\textrm{rd}}=m_{\textrm{sd}}=5$,
and $C=3$. 

\begin{figure*}[t]
\begin{centering}
\subfloat[Exact and asymptotic capacity for NOMA and OMA schemes. \textcolor{black}{Fading
parameter $m_{i}=\{1,5\}$}]{\begin{centering}
\includegraphics[scale=0.45]{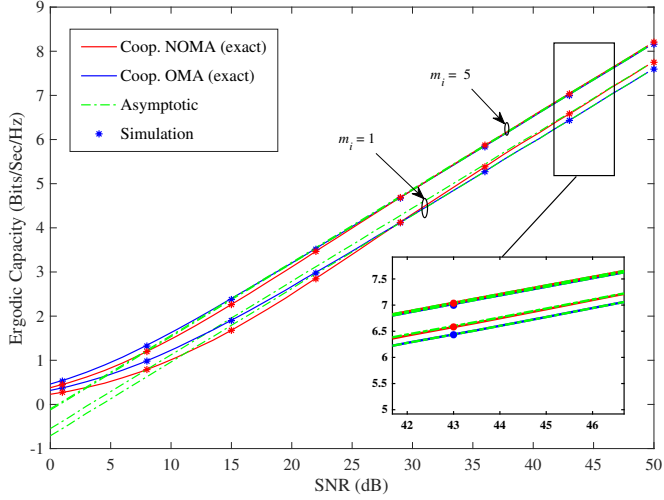}
\par\end{centering}
}\subfloat[Close-up extract of ergodic capacity for NOMA and OMA at high SNR.
One-sided Gaussian (\textcolor{black}{$m_{i}$} = 0.5), Rayleigh (\textcolor{black}{$m_{i}$}
= 1), Nakagami-$m$ (\textcolor{black}{$m_{i}$} = 2).]{\begin{centering}
\includegraphics[scale=0.45]{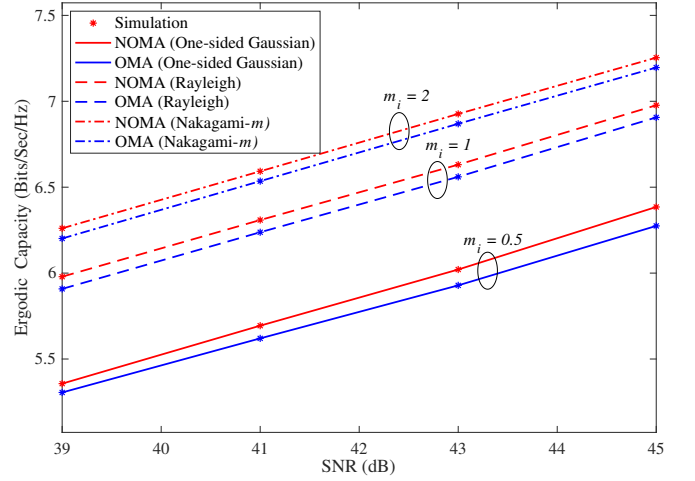}
\par\end{centering}
}
\par\end{centering}
\caption{\textcolor{black}{\label{fig:C_vs_SNR}Ergodic capacity as a function
of $\bar{\gamma}$ for both the cooperative NOMA and OMA schemes }over
the Fisher-Snedecor $\mathcal{F}$ composite fading channel,\textcolor{black}{{}
when $a_{2}=0.01$ and varying fading parameter $m_{i}$, }where $i\in\left\{ \textrm{sr},\textrm{rd},\textrm{sd}\right\} .$}
\end{figure*}

To begin with, we plot in Fig. \ref{fig:C_vs_SNR} the analytical
and simulated ergodic capacity with respect to the average SNR for
both the cooperative NOMA and OMA systems over the $\mathcal{F}$
fading channel with different multipath and shadowing conditions when
the power allocation factor $a_{2}=0.01$. It is clear from this figure
that the analytical and simulated results are in good agreement, which
verifies the accuracy of our analysis. Note that the analytical results
for the NOMA and OMA systems are obtained from (\ref{eq:Cnoma}) and
(\ref{eq:Coma2}), respectively. It is evident that the NOMA approach
outperforms the conventional one when SNR is relatively high, as shown
in the extract in Fig. \ref{fig:C_vs_SNR}b, whereas the OMA approach
tends to have better performance at low SNR values (Fig. \ref{fig:C_vs_SNR}a).
Another observation worth highlighting is that as the fading parameter
$m_{i}$ is increased, the performance of both systems are enhanced.
This is justified by the fact that increasing $m_{i}$ implies increasing
the number of multipath clusters arriving at the receiving node which
consequently improves the received SNR. This positively influences
the ergodic capacity performance. To further emphasize the versatility
of the derived expressions, we present the effect of the fading parameters
under special cases of one-sided Gaussian (\textcolor{black}{$m_{i}$}
= 0.5), Rayleigh (\textcolor{black}{$m_{i}$} = 1) and Nakagami-$m$
(\textcolor{black}{$m_{i}$} = 2) channels in Fig. \ref{fig:C_vs_SNR}b.
As can be observed, the ergodic capacity improvement is clearly evident
as the fading severity $m_{i}$ is changed. Furthermore, Fig. \ref{fig:C_vs_SNR}a
also presents the asymptotic ergodic capacity curves of the cooperative
NOMA and OMA systems; these results are obtained using (\ref{eq:NOMAasym})
and (\ref{eq:OMAasym}). It is visible that in the high SNR regime,
the asymptotic curves are in very good agreement with the exact results
and this indeed verify the accuracy of the derived asymptotic capacity
expressions. 

\begin{figure}
\begin{centering}
\includegraphics[scale=0.6]{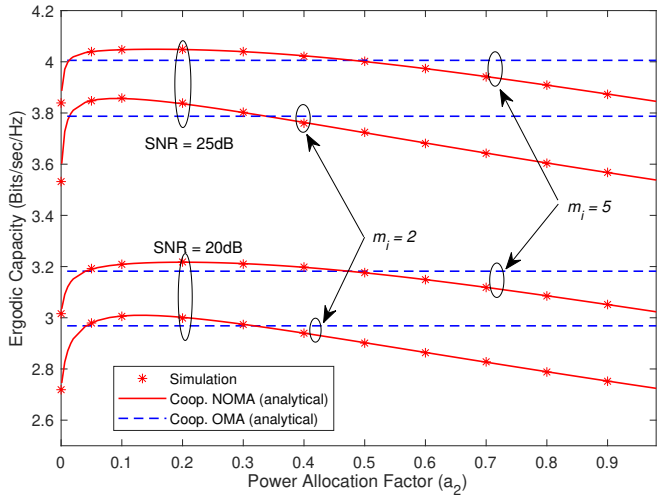}
\par\end{centering}
\caption{\textcolor{black}{\label{fig:C_vs_a2}Ergodic capacity with respect
to the power allocation factor of $R,$ $a_{2}$ and various $\bar{\gamma}$
values. The fading parameter $m_{i}=\{2,5\}$, }where $i\in\left\{ \textrm{sr},\textrm{rd},\textrm{sd}\right\} .$\textcolor{black}{{}
Results for the cooperative OMA scheme are also shown.}}
\end{figure}

\begin{figure}
\begin{centering}
\includegraphics[scale=0.6]{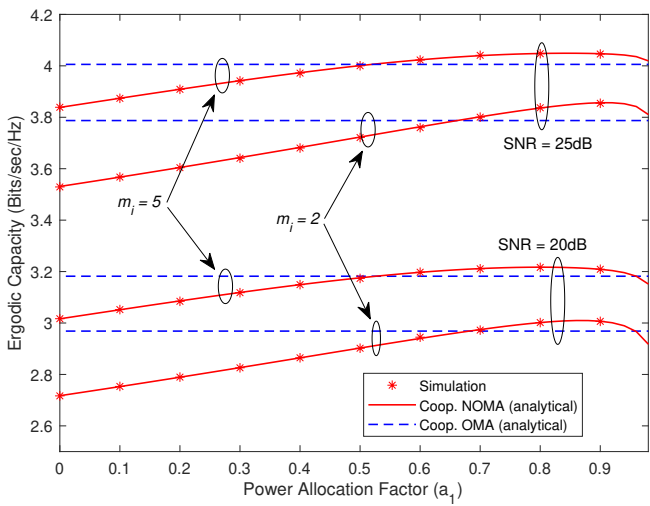}
\par\end{centering}
\caption{\textcolor{black}{\label{fig:C_vs_a1}Ergodic capacity with respect
to the power allocation factor of $D,$ $a_{1}$ and various $\bar{\gamma}$
values. The fading parameter $m_{i}=\{2,5\}$, }where $i\in\left\{ \textrm{sr},\textrm{rd},\textrm{sd}\right\} .$\textcolor{black}{{}
Results for the cooperative OMA scheme are also shown.}}
\end{figure}

To illustrate the impact of the power allocation factor on the system
performance, we present in Fig. \ref{fig:C_vs_a2} the ergodic capacity
versus the power allocation factor $a_{2}$ with different values
of $\bar{\gamma}$ and fading parameters. The performance of the cooperative
OMA system is also included in this plot for the sake of comparison.
It is interesting to see that when $a_{2}$ is either too small or
too large, the performance of the NOMA approach degrades considerably
which results in relatively poor performance compared to the OMA scheme.
It is also noted that the performance gap between the two systems
becomes more pronounced when $\textrm{SNR}=25\textrm{dB}$ compared
to the case when $\textrm{SNR}=20\textrm{dB}$. Similarly, in Fig.
\ref{fig:C_vs_a1}, we plot the ergodic capacity versus the power
allocation factor $a_{1}$ with different values of $\bar{\gamma}$
and fading parameters. Again, it can be observed that the performance
of the NOMA scheme varies considerably when $a_{1}$ is either too
small or too large. This performance mirrors the change in $a_{2}$
as expected since $a_{1}+a_{2}=1.$ Furthermore, in Figs. \ref{fig:C_vs_a2}
and \ref{fig:C_vs_a1}, the presence of a maxima indicates that when
the power allocation factor is carefully selected, the NOMA approach
is able to offer better performance, which means that optimizing this
factor is crucial to maximizing the performance of the NOMA system.

\begin{figure}
\begin{centering}
\includegraphics[scale=0.45]{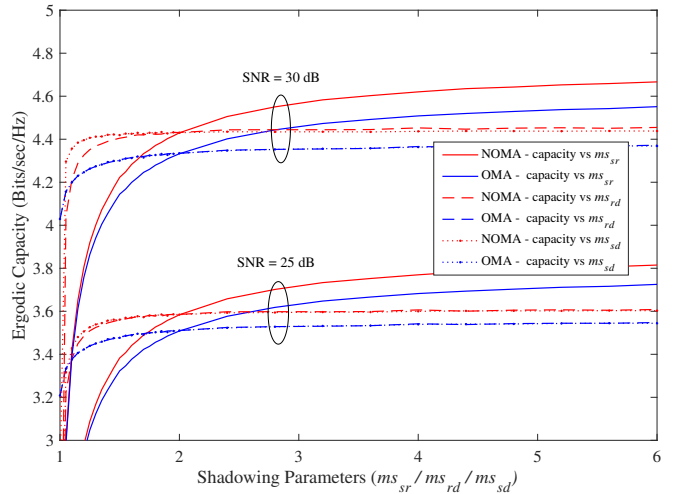}
\par\end{centering}
\caption{\textcolor{black}{\label{fig:C_vs_ms}Ergodic capacity of NOMA and
OMA schemes as a function of the shadowing parameters of }S-to-R ($m_{s_{\textrm{sr}}}$),
R-to-D ($m_{s_{\textrm{rd}}}$) and S-to-D ($m_{s_{\textrm{sd}}}$)
links\textcolor{black}{{} for various $\bar{\gamma}$ values.}}
\end{figure}

In Fig. \ref{fig:C_vs_ms}, we present the ergodic capacity of both
NOMA and OMA schemes as a function of the shadowing parameters of
the various links. Here we assume the fading parameters $m_{\textrm{sr}}=m_{\textrm{rd}}=m_{\textrm{sd}}=2$
and $C=2$. Also, while plotting the curves for the shadowing parameters
on one link, we assume a fixed shadowing parameter for the other 2
links (moderate shadowing $m_{s_{i}}=4$). We first observe that for
regions with severe shadowing $m_{s}<2$, the NOMA S-to-D link provides
the best ergodic capacity, while the NOMA R-to-D link rapidly becomes
improved to a similar capacity. Further decrease in the shadowing
severity (as $m_{s}\longrightarrow\infty)$ presents noticeable improvements.
On the other hand, for the OMA scheme, the capacities of both the
S-to-D and R-to-D links are fairly even (and lower) through all the
regions. The best capacity improvements for the system can however
be achieved when the shadowing severity becomes moderate to low i.e.
$m_{s}>2$, because for the S-to-R link, both the NOMA and OMA schemes
perform much better. This performance increase can be attributed to
the configuration of the system with power allocation factor of 90\%
in favor of this link. Therefore, this further indicates the importance
of selecting an appropriate power allocation factor. 
\begin{figure}
\begin{centering}
\includegraphics[scale=0.6]{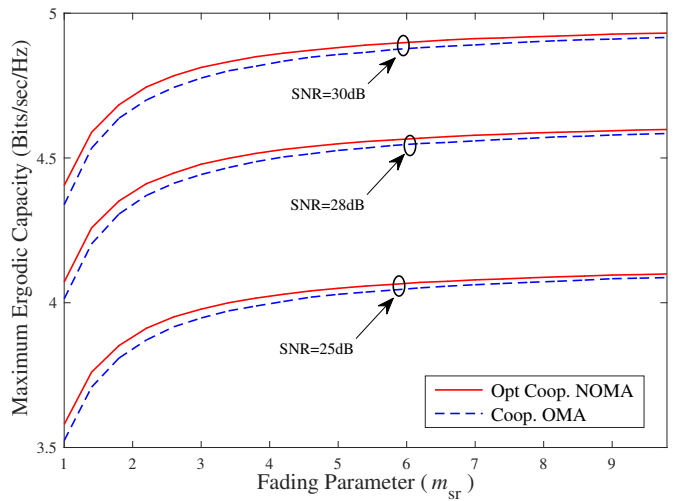}
\par\end{centering}
\caption{\textcolor{black}{\label{fig:C_vs_xx}Maximum achievable ergodic capacity
of the NOMA system over }the Fisher-Snedecor $\mathcal{F}$ composite
fading channel\textcolor{black}{{} versus $m_{\textrm{sr}}$ for different
values of $\bar{\gamma}$. Note that it is assumed here that $m_{\textrm{rd}}=m_{\textrm{sd}}=5.$}}
\end{figure}

Next, we investigate the impact of optimizing the power allocation
parameter $a_{2}$ on the ergodic capacity performance. It should
be mentioned that due to the complexity of the derived expression
in (\ref{eq:Cnoma}), it is not possible to obtain the optimal $a_{2}$
in closed-form. However, it does not pose any difficulty to obtain
numerical solutions using software tools. In this respect, extensive
simulations were conducted to find the maximum achievable ergodic
capacity that corresponds to the optimal $a_{2}$. Fig. \ref{fig:C_vs_xx}
depicts the maximum achievable ergodic capacity versus the fading
parameter $m_{\textrm{sr}}$ for $\bar{\gamma}=25\textrm{dB},\,28\textrm{dB}\,\textrm{and}\,30\textrm{dB}$,
when \textcolor{black}{$m_{\textrm{rd}}=m_{\textrm{sd}}=5$.} One
can clearly see from these results that the optimized NOMA system
always outperforms the conventional one for all the system configurations
under study. It is also worthwhile pointing out that increasing the
fading parameter of the S-to-R link will improve the capacity for
both cooperative NOMA and OMA systems. 

\section{\textcolor{black}{\label{sec:Conclusion}Conclusion }}

This paper has been devoted to the analysis of cooperative relaying
wireless networks based on NOMA over Fisher-Snedecor $\mathcal{F}$
composite fading channels. The performance of conventional cooperative
relaying systems, i.e., based on OMA, has also been studied for the
sake of completeness and comparison. For both the NOMA and OMA systems
under consideration, we have derived exact and asymptotic closed-form
expressions of the ergodic capacity\textcolor{black}{{} in terms of
special functions and converging power series}. These expressions
have been used to investigate the impact of various system and fading
parameters on the capacity performance. Results have shown that the
NOMA approach is able to achieve better performance compared to the
OMA scheme in the high SNR region. Results have also demonstrated
the positive impact of the multipath fading and the shadowing parameters
on the system performance. In addition, optimizing the power allocation
factor in the NOMA system is crucial to maximizing the average capacity. 

\bibliographystyle{ieeetr}
\bibliography{bib}

\end{document}